\documentclass[fleqn,usenatbib]{mnras}

\usepackage{newtxtext,newtxmath}
\usepackage[T1]{fontenc}
\usepackage{ae,aecompl}

\usepackage{graphicx}
\usepackage{amsmath}
\usepackage{amssymb}
\usepackage{adjustbox}

\usepackage{epstopdf}
\usepackage{enumerate}
\usepackage{wrapfig}
\usepackage{lscape}
\usepackage{rotating}
\usepackage{color}

\setcitestyle{citesep={,}}

\title[CEMP stars in APOGEE DR12]{Carbon-enhanced metal-poor stars in the SDSS-APOGEE database}

\author[C. L. Kielty et al.]{Collin L. Kielty,$^{1}$ \thanks{E-mail: clkielty@uvic.ca}
K. A. Venn,$^{1}$
N. B. Loewen,$^{1}$
M. D. Shetrone,$^{2}$
\newauthor
V. M. Placco,$^{3}$ 
F. Jahandar,$^{1}$ 
Sz. M{\'e}sz{\'a}ros,$^{4,5}$
and S. L. Martell$^{6}$
\\
$^{1}$Department of Physics and Astronomy, University of Victoria, Victoria, British Columbia, V8W 3P2, Canada\\
$^{2}$McDonald Observatory, University of Texas at Austin, HC75 Box 1337-MCD, Fort Davis,TX 79734\\
$^{3}$Department of Physics and JINA Center for the Evolution of the Elements, University of Notre Dame, Notre Dame, IN 46556, USA\\
$^{4}$ELTE Gothard Astrophysical Observatory, H-9704 Szombathely, Szent Imre Herceg st. 112, Hungary \\
$^{5}$Premium Postdoctoral Fellow of the Hungarian Academy of Sciences \\
$^{6}$School of Physics, University of New South Wales, Sydney, NSW 2052, Australia
}

\date{Accepted XXX. Received YYY; in original form ZZZ}

\pubyear{2017}

\begin{document}
\label{firstpage}
\pagerange{\pageref{firstpage}--\pageref{lastpage}}
\maketitle

\begin{abstract}
We identify six new CEMP stars ([C/Fe]$> +0.7$ and [Fe/H]$< -1.8$) and another seven likely candidates within the APOGEE database following Data Release 12. These stars have chemical compositions typical of metal-poor halo stars, e.g., mean [$\alpha$/Fe] = $+0.24\pm0.24$, 
based on the ASPCAP pipeline results.  A lack of heavy element spectral lines impedes further sub-classification of these CEMP stars, however, based on radial velocity scatter, we predict most are not CEMP-$s$ stars which are typically found in binary systems. Only one object, 2M15312547+4220551, may be in a binary since it exhibits a scatter in its radial velocity of $1.7 \pm 0.6$ km s$^{-1}$ based on three visits over a 25.98 day baseline. Optical observations are now necessary to confirm the stellar parameters and low metallicities of these stars, to determine the heavy-element abundance ratios and improve the precision in the derived abundances, and to examine their CEMP sub-classifications.
\end{abstract}

\begin{keywords}
surveys -- stars: abundances -- stars: carbon -- stars: chemically peculiar -- Galaxy: halo
\end{keywords}

\section{Introduction}\label{sIntro}

The arrival of large multi-object spectroscopic surveys  in the past decade has accelerated the fields of stellar archaeology and near-field cosmology by providing homogeneous and precise stellar parameters and abundances for $\sim10^5$ stars in all structural components of the Galaxy. With such large data samples and using stellar chemical abundance profiling, it is possible to probe the primary astrophysical processes responsible for early star formation, and gain insight into Galactic formation and evolution.

An early endeavour into large scale spectroscopic surveys includes the RAdial Velocity Experiment (RAVE; \citet{Steinmetz06}). RAVE collected $R \sim 7000$ spectra, measured radial velocities and proper motions accurate to 1.5 km/s using the Calcium triplet, and determined stellar parameters and elemental abundances (Mg, Al, Si, Ca, Ti, Fe, and Ni) for $\sim 480,000$ stars with $8 < I < 12$ \citep{Boeche11}. From this dataset, new constraints have been placed on the Galactic mass and escape velocity \citep{Smith07, Piffl14a, Piffl14b}, the Aquarius tidal stream has been discovered \citep{Williams11, WdB12}, tidal debris around globular clusters have been identified and characterized \citep{Kunder14, Anguiano15, FT15}, and a plethora of studies on Galactic disc kinematics and chemical gradients have been carried out \citep{Ruchti10, Ruchti11, Wilson11, Boeche13a, Boeche13b, Williams13, Binney14, Boeche14}.

Exploring down to $g\geq20$ magnitude, the SDSS/SEGUE survey \citep{Abazajian09, Yanny09a} collected $R \sim 2000$ optical (3850-9200 \AA) spectra for $ \sim 300,000 $ stars with the intent of mapping the kinematics and stellar populations of the Milky Way. The depth of the SEGUE survey has allowed for the kinematic characterization of the Galaxy \citep{Smith09, Carollo10, Bond10, Gomez12, Bovy12a, Bovy12b}, the discovery and characterization of faint substructures within the Galaxy, namely the Orphan and Sagittarius streams as well as the Segue 1 and 2 satellites, \citep{Belokurov07, Klement09, Yanny09b, Belokurov09, Newberg10}, and the investigation of chemistry within these structures, including the discovery of chemically peculiar stars \citep{An09, Norris10a, Norris10b, Martell10, Aoki10, Simon11, Lee11, Lee13, Schlesinger12, Santucci15, Lee17}. Modern surveys such as Gaia-ESO \citep{Gilmore12}, GALAH \citep{DeSilva15}, and APOGEE \citep{Majewski15} will critically increase the quality and depth of our understanding of the Galaxy with higher resolution, larger sample sizes, and high precision measurements.

Complimentary to these large spectroscopic surveys are more targeted surveys of metal-poor stars. \citet{McWilliam95}, \citet{Aoki07}, \citet{Yong13}, \citet{Norris13}, and \citet{Lee13} which have shown the presence of a large metal-poor population in the stellar halo. Within this population of metal-poor stars, a significant fraction of stars with abundance anomalies have been found. Of particular interest in the chemically peculiar star group are those with carbon-enhancement \citep[CEMP stars;][]{BC05} which represent 20\%\ of stars with [Fe/H]$ < -2.0 $ and with a rapidly increasing fraction at lower metallicities, approaching unity for known stars with [Fe/H] $< -4.5$ \citep{Christlieb03, Lucatello05, Frebel06, Carollo12, Aoki13, Lee13, Norris13, Yong13, Placco14, Hansen16no}.

CEMP stars have been the focus of a large number of recent studies due to their importance in identifying rare processes in the context of Galactic chemical evolution. Within the population of CEMP stars several subclasses exist, defined by ratios of neutron-capture elements in the stellar spectra: CEMP-$s$, CEMP-$r$, CEMP-$r/s$ (or CEMP-$i$), and CEMP-no. Each of these subtypes are described below.

The CEMP-$s$ stars (those with slow neutron-capture, $s$-process, element enhancement) are proposed to be the metal-poor analogues to the \ion{Ba}{ii}, classical CH, and subgiant CH stars, based on similar abundance patterns \citep{PS01, Sneden03}. Likewise, CH stars demonstrate a binary frequency "consistent with unity" \citep{McCW90}, an observational trend shared by CEMP-$s$ stars as shown by \citet{Lucatello05}, \citet{Starkenburg14}, and \citet{Hansen16s}. The peculiar abundance patterns seen in CH stars, combined with the high observed binary fraction are indicative of the accretion of material from an intermediate-mass AGB companion, either through Roche lobe overflow or through efficient stellar winds \citep{Han95}. The latter scenario is more likely as a result of the instability of the Roche lobe in thermally pulsing AGB stars and the typically larger spatial separations of these binary systems \citep{Paczynski65, Abate13}. Similarly, \citet{Herwig05}, \citet{Placco13}, and \citet{Hansen16a} have shown the abundance profiles of the CEMP-$s$ stars are consistent with enrichment from an AGB companion. The difference between these similarly natured objects currently exists only in an arbitrary cut in metallicity.  An upper limit of [Fe/H]$<-1.8$ is described by \citet{Lucatello05} to separate CEMP-$s$ stars from the more metal-rich classical \ion{Ba}{ii}, CH, and subgiant CH stars. The most metal-poor CH stars have been observed down to [Fe/H]$\sim -1.5$  \citep{Vanture92, Goswami05}, and the \citet{Lucatello05} cut seeks to establish a factor of two difference in metallicity between these systems.

CEMP-$r/s$ stars show an enhancement of rapid neutron-capture, $r$-process, elements as well as elements from the $s$-process. The origins of these stars are currently under investigation with theories ranging from the formation of a CEMP-$s$ star via AGB companion mass transfer in an environment previously enriched with $r$-process materials \citep{Jonsell06} to stars influenced by the intermediate neutron-capture, $i$-process, which may occur in a range of stellar sites \citep{Roederer14, Dardelet15, Hampel16}.

CEMP-no stars (those with no $n$-capture enhancements) are potentially the most informative in the context of Galactic chemical evolution since they do not appear to be closely linked with binary systems, opening doors to other carbon-enhancement mechanisms beyond mass transfer. \citet{Maynet06} and \citet{Maeder15} have proposed fast rotating metal-poor "spinstars" experience partial mixing processes that bring CNO materials to the stellar surface. Through stellar winds, their local ISM becomes enriched with these elements and the later generations of stars to form in these regions would exhibit CEMP abundance profiles. C-enhancement of the ISM via Population III faint supernovae has also been modelled in detail by \citet{UN03}, \citet{UTINM06}, \citet{HegerWoosley10}, and \citet{Tominaga14}, who have shown that fined tuned levels of mixing and fallback during the supernova can result in abundance profiles consistent with those seen in the CEMP-no stars.  Regardless of the exact mechanism(s) responsible for the abundances seen in CEMP-no stars, it appears these old objects may reflect the nucleosynthetic enrichment processes present in the early Universe. 

Recent studies have shown that distribution of absolute C abundance $A$(C)$=\log \epsilon$(C) for CEMP stars splits into at least two distinct `bands` based on their evolutionary history \citep{Spite13, Bonifacio15, Hansen15a, Yoon16}. \citet{Yoon16} identifies peaks in the $A$(C) distribution at $A$(C)=7.96 and $A$(C)=6.28, corresponding to the high-C and low-C regions respectively. They argue the separation of these two bands serves as an effective and astrophysically motivated metric in assessing the history, nature, and sub-class of CEMP stars, as the vast majority of known CEMP-$s$ and CEMP-$r/s$ stars are highly concentrated around the high-C band while known CEMP-no stars are scattered around the low-C band. This separation allows for a preliminary classification of CEMP stars based solely on their $A$(C), rather than on the abundance ratios of neutron-capture elements. \citet{Yoon16} successfully classified 87\% (139 of 159) of the CEMP-$s$ and CEMP-$r/s$ stars and 93\% (104/112) of the CEMP-no in their sample using only $A$(C), and treating the traditional [Ba/Fe] criterion as the standard. Additionally, \citet{Yoon16} observe the known and likely binaries in their sample separate around the midpoint in the distribution at $A$(C) = 7.1, further supporting extrinsic origins of carbon enhancement in the CEMP-$s$ and CEMP-$r/s$ stars and intrinsic origins for CEMP-no stars. An understanding of the CEMP stars may prove critical in unlocking knowledge on early Galactic astrophysical processes, pristine stellar populations, and would assist in completing the picture of Galactic evolution.

CEMP stars have been found serendipitously in spectroscopic surveys. Often, low resolution spectroscopy focused on the carbon sensitive \textit{G}-band and metallicity sensitive Ca II K feature in the optical is used to identify CEMP candidates (see the Hamburg/ESO (HES) survey \citep{Christlieb01, Rossi05, Placco10, Placco11}). High resolution spectroscopic follow up is then used as confirmation on the nature of these stars. 

Playing off large sample size statistics, the APOGEE database may serve as a useful tool in the search for these rare objects and provide us with new candidates for follow-up optical spectra. In this paper we explore the APOGEE database for new CEMP stars. In Section 2 we summarize the key elements of the APOGEE survey and the selection of our CEMP candidates, Section 3 explores the ASPCAP abundances of our candidates in detail, independent abundances are derived and cross checked in Section 4, and our discussion, concluding remarks and perspectives are gathered  in sections 5 and 6.

\section{The APOGEE Spectroscopic Survey}\label{sAPOGEE}

The APOGEE survey of the the Sloan Digital Sky Survey, SDSS-III Data Release 12 \citep{Eisenstein11, Alam15} provides high resolution ($R \backsim 22,500$) IR (H-band) spectra for $\backsim$150,000 targets and derives chemical abundances for 15 elements: C, N, O, Na, Mg, Al, Si, S, K, Ca, Ti, V, Mn, Fe, and Ni using the APOGEE Stellar Parameters and Chemical Abundances Pipeline \citep[ASPCAP;][]{GarciaPerez15}. Recent studies such as the colour-$T_{\rm eff}$-[M/H] relationship for cool dwarfs \citep{Schmidt16}, the separation, in chemical abundance space, of stellar populations and structures within the Galaxy \citep{Bovy16b}, and the discovery of halo stars with globular cluster origins via chemical tagging \citep{Martell16}, are only now probing the surface of this rich dataset's resourcefulness. Halo fields comprise $\sim$25\% of the total sample \citep{Zasowski13} making APOGEE a great resource to map chemical abundances in the metal-poor regions of the Galaxy and to search for previously unobserved CEMP stars. 

Since the classification of VMP and CEMP stars is dependent solely on metallicity and [C/Fe], it is important to address the distinction between [M/H] and [Fe/H] in APOGEE. While ASPCAP can return [Fe/H] for an object, the primary derived metallicity is given as [M/H]. This overall metal abundance is scaled to the solar abundance pattern through spectral template fitting, and is derived by tracking all metals over the entire wavelength regime of the APOGEE spectrograph. By comparing the derived [M/H] metallicity of well studied clusters to the spectroscopic metallicity found in the literature, the metallicity scale was calibrated to a [Fe/H] scale \citep{Meszaros13, Zamora15, Holtzman15}.

 \citet{Holtzman15} found that a difference of $\sim$0.2 dex between the calibrated [Fe/H] and [M/H] is seen below [M/H]=$-1.0$, with the discrepancy increasing as a function of decreasing metallicity, reaching $\simeq$ 0.3 dex around [M/H]$=-2.0$ (see their Figure 6). The ASPCAP output is limited to metallicities [M/H]$> -2.4$ as the Fe lines become too weak to measure at lower metallicity in infrared spectra. Stars more metal-poor than this could still be present in the survey, but  ASPCAP would either return [M/H] $= -2.4$ (a proxy for an upper limit); alternatively, it may report the star as  hotter, since RGB spectral lines would be weaker, or chemically peculiar. This restriction on metallicity, coupled with the higher associated errors in the metal-poor regime, have motivated a majority of the previous APOGEE-based studies to restrict the analysis to near-solar metallicity stars. This paper serves as a step into the metal-poor regime.

\subsection{Uncertainties in the ASPCAP Abundances }\label{sUncert}

The APOGEE team has identified issues driving uncertainty in the ASPCAP stellar parameters and abundances. Other than the known persistence problem (see Section \ref{sSample}), the spectral quality itself is not a significant source of uncertainty since APOGEE targets are reobserved multiple times until a $SNR > 100$ is attained in the combined spectrum for each object. A complete discussion of the calibration process and uncertainties can be found in \citet{Meszaros13}, \citet{Nidever15}, and \citet{Holtzman15}, and we highlight the issues relevant to metal-poor stars below.

The primary metallicity uncertainty in the ASPCAP output is found at low metallicity, which can be seen in the data calibration to open and globular clusters. \citet{Holtzman15} used literature data on 20 open and globular clusters to calibrate $T_{\rm eff}, \log g,$ metallicity, $\alpha$, C and N  for the APOGEE DR12 data. For [M/H] metallicities $< -1.0$, the difficulty of detecting Fe lines in IR spectra results in systematic differences between the ASPCAP results and the literature of up to 0.2 dex (see their Figure 6). This effect additionally increases with temperature as spectral lines weaken. 

\citet{Meszaros13} also examined the large errors in abundances in metal-poor stars due to the impact of molecular bands (C, N, and O particularly) in the metal-poor stars. These errors stem from the reduced number of spectral features at low metallicity and a relatively strong dependence of the line strength on the stellar parameters, particularly $T_{\rm eff}$. Additionally, a calibration of ASPCAP derived [C/Fe], [N/Fe] and [O/Fe] to literature values for globular cluster stars cannot be generally applied, since globular clusters have prominent star-to-star variations in light element abundances that are not common in the field \citep{Holtzman15}. 

CEMP stars present an intriguing challenge for ASPCAP. A CEMP star with [Fe/H]$<-2.5$ and [C/Fe]$>+1.0$ is near the limits of the APOGEE synthetic spectral library, and so the best-fit values for the stellar parameters can be rather different than if the analysis pipeline were able to further adjust the abundances of Fe and C. As an example, the derived $T_{\rm eff}$ may be artificially high to reproduce weak Fe lines in the observed spectrum when the true [Fe/H] is lower than allowed by the synthetic library.

\begin{figure*}
\centering
\includegraphics[width=\textwidth]{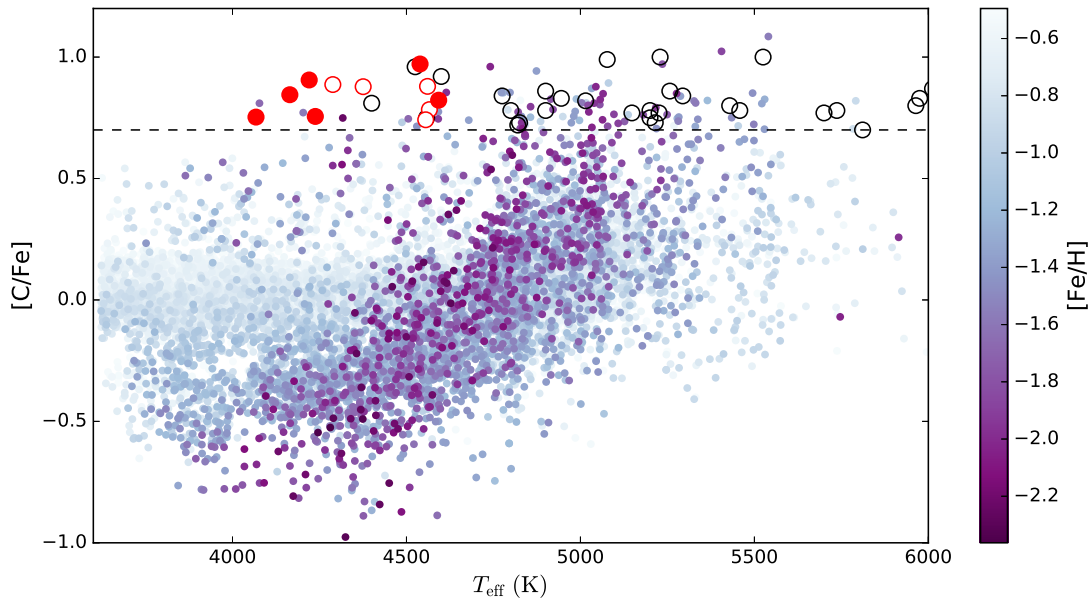}
\caption{The relationship between [C/Fe] and $T_{\rm eff}$ for all metal-poor ([Fe/H$< -0.5$) APOGEE stars, coloured by metallicity. Filled red circles are our Group A CEMP candidates, open red circles are Group B (see Section \ref{sSample} for the distinction between Group A and B) and open black circles are known CEMP stars selected from \citet{Placco14}. Stars with [Fe/H]$<-1.5$ demonstrate a very clear trend between [C/Fe] and $T_{\rm eff}$ suggesting a correlation between [C/Fe] and $T_{\rm eff}$. This relationship is non-physical and likely the combined result of the double-metal CO molecular bands' temperature sensitivity and the inclusion of upper-limit abundances derived from weak lines.}
\label{CvT}
\end{figure*}

\begin{figure}
\centering
\includegraphics[width=\linewidth]{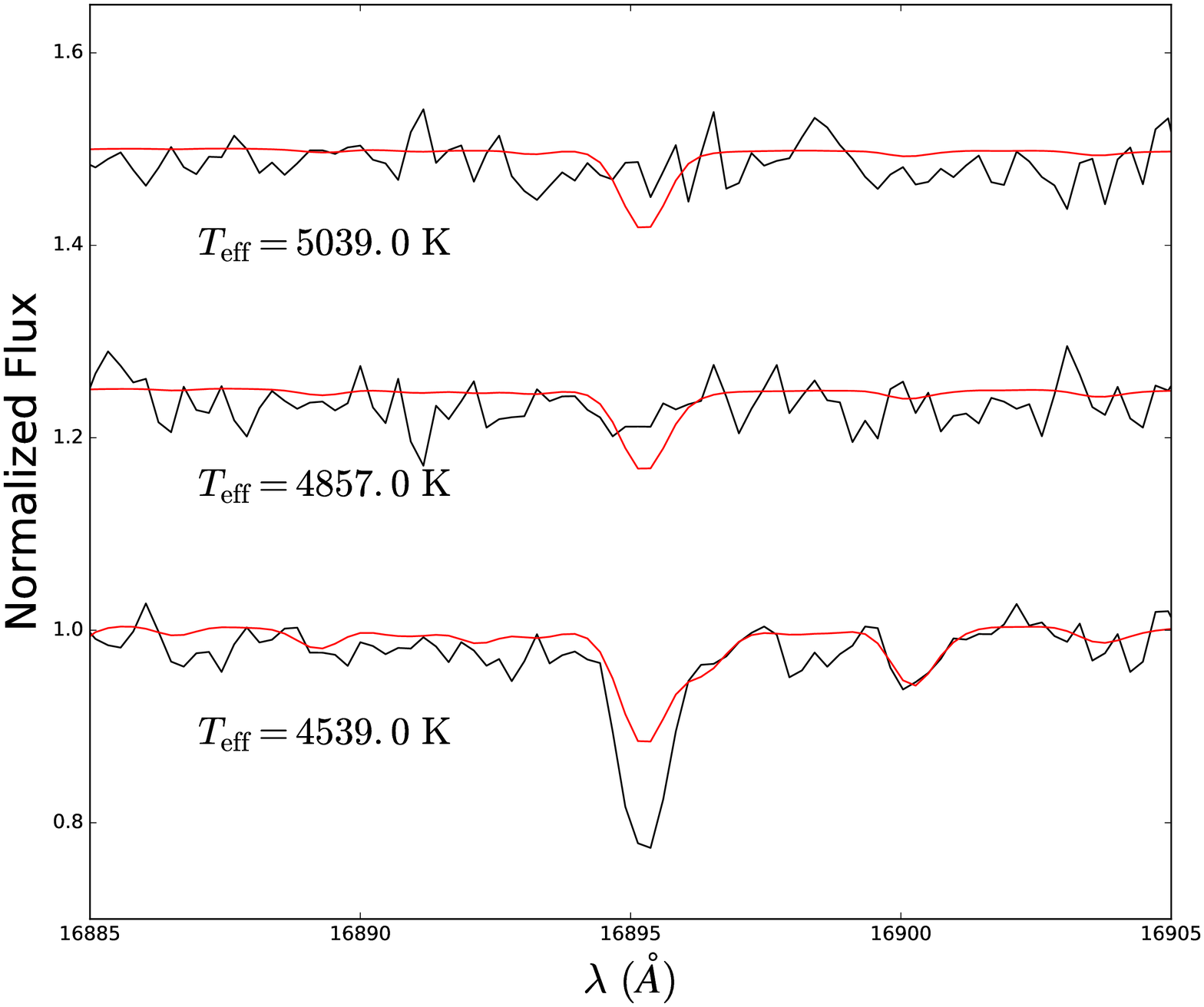}
\caption{Sample APOGEE combined spectra (black) and ASPCAP synthetic spectra (red), centred on the atomic \ion{C}{i} line, for three stars (2M21295801+1214260, 2M14442263+1350570, and 2M16300629-1252459; top to bottom) with [Fe/H]$\sim -1.95$, [C/Fe]$\sim +0.8$, and $S/N \sim 100$, but varying $T_{\rm eff}$. The atomic \ion{C}{i} line is not visible for metal-poor stars with $T_{\rm eff} > 4600$K indicating the reported C abundance by APOGEE should be treated as an upper-limit for these warmer objects. The poor synthetic fit of the atomic \ion{C}{i} line in 2M16300629-1252459 suggests a higher than reported C abundance for this star.}
\label{Upperlims}
\end{figure}

In Figure \ref{CvT}, we show the ASPCAP [C/Fe] vs. $T_{\rm eff}$ and [Fe/H]. Stars with metallicities near solar ([Fe/H]$\gtrsim -0.5$) are well clustered around solar [C/Fe] and do not display any temperature dependence beyond a higher degree of scatter for $T_{\rm eff} \gtrsim 4700$ K, as expected based on the the aforementioned uncertainties in the spectral analysis. The lower metallicity stars ([Fe/H] $<-1.5$), however, show a very clear trend between [C/Fe] and $T_{\rm eff}$, indicating that carbon abundances are unreliable in metal-poor giants. This relationship is the combined result of the temperature sensitivity of CO molecular bands, continuum placement effects, and upper-limit abundances derived from weak lines. Upper-limit abundances are not identified or flagged by ASPCAP, requiring one to examine the ASPCAP spectrum and synthetic fit manually. The issue of upper-limits for the warmer stars is displayed in Figure \ref{Upperlims}. Three stars with similar [C/Fe] and [Fe/H], but varying $T_{\rm eff}$ were selected from APOGEE to examine the relationship between the atomic \ion{C}{i} line strength and $T_{\rm eff}$. The atomic \ion{C}{i} line at 16895 \AA\ provides one of the most reliable estimates for the C abundance in the $H$-band (see Section \ref{sMOOGCNO}), and thus the quality of the model fit to this feature is a critical test of the C abundance. Despite similar [Fe/H], [C/Fe], and $S/N$, the atomic \ion{C}{i} line becomes indistinguishable from the noise in the spectra of stars with $T_{\rm eff} \gtrsim 4800$ K. Consequently, C abundances derived for warmer stars should be treated as upper limits, diminishing the likelihood these stars are actually carbon-enhanced. The cooler, C-poor, metal-poor stars in Figure \ref{CvT} may be subject to similar systematic bias. In summary, low metallicity stars at both high $T_{\rm eff}$ - high [C/Fe] and low $T_{\rm eff}$ - low [C/Fe] in the APOGEE DR12 database are subject to systematics that have not been accounted for and should be handled with caution.

Previous CEMP studies often impose a selection of  $T_{\rm eff} > 4800$ K, as well as $\log g \geq 1.3$. This minimizes the possibility the selected stellar sample is contaminated by cooler AGB stars, which can have very similar surface abundance profiles to CEMP stars as the result of the third dredge-up \citep{Lucatello05}. The inclusion of previously known CEMP stars (selected from \citet{Placco14}) in Figure \ref{CvT} highlight this selection of CEMP stars at higher $T_{\rm eff}$, indicating that the DR12 ASPCAP abundances are not ideal for detecting the warmer CEMP stars that are typically studied in the literature, and may be biased towards cooler AGB contaminants.

\subsection{Selected Sample from the APOGEE Database}\label{sSample}

Our sample in Figure \ref{CvT} was collected from the SDSS Sky Server by querying all APOGEE objects with no bad data flags. Guided by the offset between [M/H] and [Fe/H] at low metallicity, and for transparency with previous studies, we adopt the calibrated [Fe/H] (\texttt{aspcapStar.fe\_h}) rather than [M/H] for the remainder of the discussion. We impose an upper limit of [Fe/H]$<-1.8$ to our sample following \citet{Lucatello05}, reducing the complete APOGEE to 425 metal-poor candidates.

In DR12, the reported abundances for a particular species $X$ is given as the logarithmic ratio [$X$/H]. Again for transparency, [$X$/Fe] abundances were calculated and the errors on [$X$/H] and [Fe/H] added in quadrature to estimate $\sigma$[$X$/Fe]. Without exception and following similar arguments for Fe, [C/Fe] was calculated from [C/H] and [Fe/H], rather than directly adopting [C/M].  Applying the definition of carbon-enhancement as [C/Fe] $>= +0.7$ \citep{Aoki07}, the sample was further reduced to 37 CEMP candidates.

Motivated by the observed trend in [C/Fe] vs. $T_{\rm eff}$ at low metallicity and the presence of the atomic \ion{C}{i} line in the lower temperature stars, we reject the suggested \citet{Lucatello05} selection cuts of  $T_{\rm eff} \geq 4800$K and $\log g \geq 1.3$ and only select stars with $T_{\rm eff} \leq 4600$K and no cut in $\log g$, reducing the sample to 13 stars. These criteria were initially adopted by \citet{Lucatello05} to minimize AGB contamination in their sample, thus the dismissal of these cuts warrants careful analysis to exclude the selection of AGB stars and is addressed in Section \ref{sCN} and \ref{sCEMPConfirmed}. The catalogue of CEMP stars by \citet{Placco14} contains 15 stars with $T_{\rm eff} < 4800$K and $\log g < 1.3$, indicating CEMP stars with these uncommon stellar parameters have been previously identified.

Amongst the 13 new CEMP candidates, the spectral quality varies significantly. \citet{Majewski15} and \citet{Nidever15} have identified distortions in the APOGEE spectra as a result of a persistence effect where the latent charge from a previous exposure remains on the CCD chip. This results in an artificially raised level and is not consistent across all three of the APOGEE chips. Super-persistence affects only the "blue" and "green" detectors and occurs more frequently in the individual visits of fainter targets (we refer the readers to \citet{Nidever15} for a more thorough discussion). Characterization of persistence is complicated and no attempt to correct the issue was implemented in DR12. 

\begin{figure}
\centering
\includegraphics[width=\linewidth]{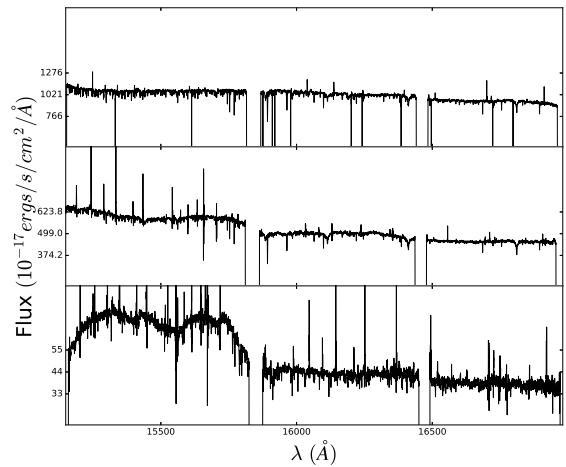}
\caption{Sample APOGEE combined spectra for three objects in our sample. The upper plot shows a Group A spectrum (2M02121851+4923143)  with no apparent persistence issues, middle shows a Group B spectrum (2M14010561+2820306) with slight persistence in the blue chip, and the bottom shows a Group C spectrum (2M12473823-0814340) with strong persistence.}
\label{Persistence}
\end{figure}

\begin{figure}
\centering
\includegraphics[width=\linewidth]{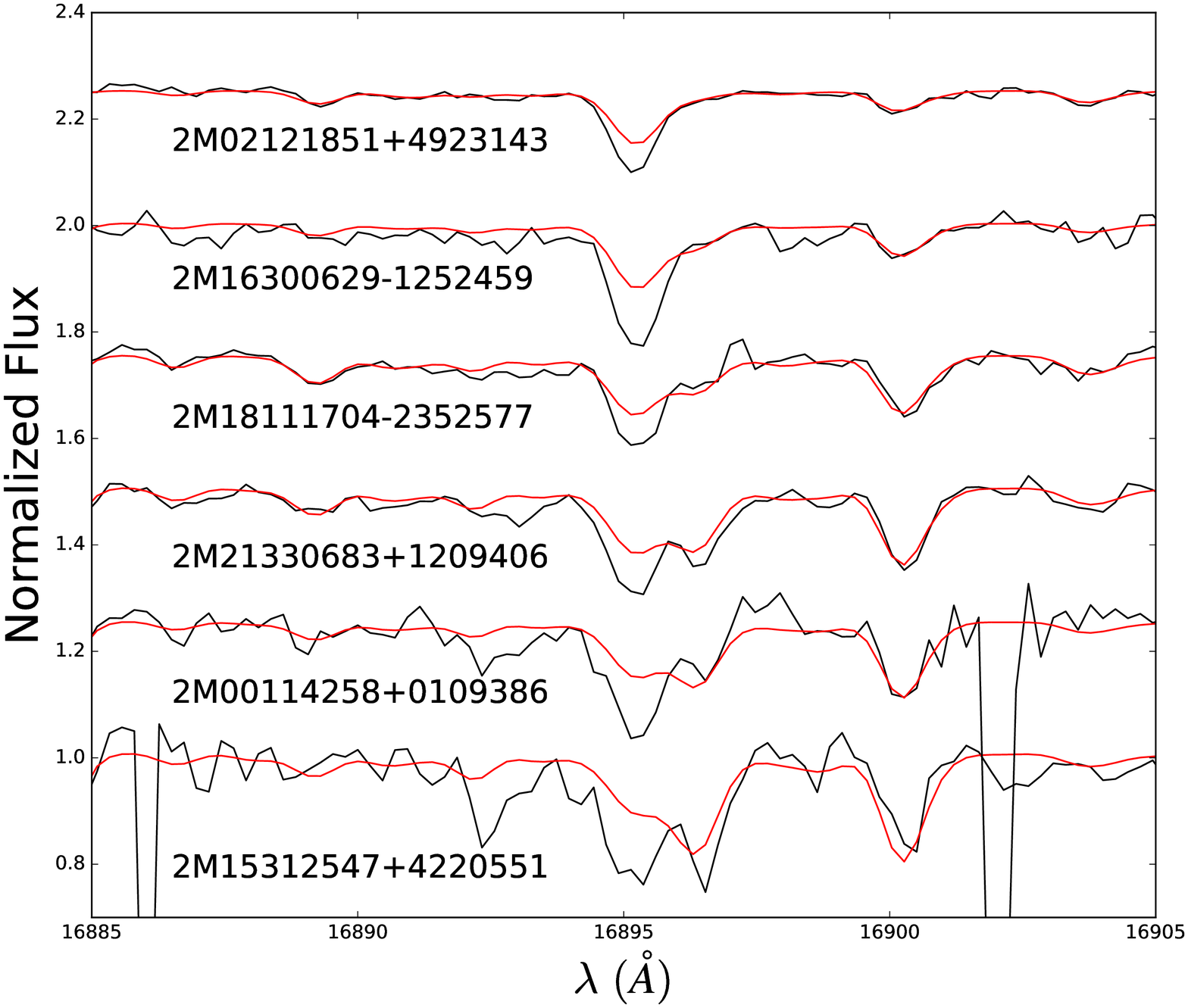}
\caption{The ASPCAP normalized spectra (black) and ASPCAP synthetic spectral fit (red) for the six Group A stars centred around the atomic \ion{C}{i} line at 16895 \AA. We observe the synthetic fit for all six stars does not accurately reproduce the atomic line suggesting the carbon abundance is higher than reported by APOGEE for these stars}
\label{GroupASpec}
\end{figure}

With these systematics in mind, we have carefully examined the individual lines and synthetic spectra from APOGEE in the database for our 13 candidates. Furthermore, DR13 included an attempt to lower the weight of individual visits when persistence is detected. While DR13 is an improvement, careful comparisons with published stars show this issue is not completely resolved (e.g. Jahandar et al. in prep.) Due to the very strong signal left over from persistence, stars which suffered from this issue were easily identified. Figure \ref{Persistence} highlights this effect, showing sample spectra for stars with no apparent persistence issues, slight persistence and strong persistence issues. Based on these spectral tests, we further separate our sample of 13 CEMP candidates into three subgroups:
\begin{itemize}
\item \textit{Group C}: Two stars were identified whose synthetic spectra poorly matched the combined spectra from APOGEE, identified as Group C in Table \ref{AllStars}. For this group, persistence or flat fielding issues were a significant issue in the combined spectra and as a result, it is highly unlikely that the stellar parameters and abundances returned by ASPCAP are reliable. These stars will be removed from further discussion.

\item \textit{Group B}: Five stars have well-matched synthetic spectra that are sufficiently consistent across the three chips which suggests these are CEMP stars. However, the combined spectra show noise levels, largely in the form of persistence on the blue chip, high enough to raise questions on the reliability of the ASPCAP results. A fresh analysis after removing the spectra affected by persistence is needed for a more precise analysis of these stars. These are the Group B stars in Table \ref{AllStars}. 

\item \textit{Group A}: The remaining six stars have excellent data, unaffected by persistence. We expect the stellar parameters and ASPCAP chemical analysis to be reliable and thus, we adopt the calibrated stellar parameters $T_{\rm eff}$, $\log g$, [Fe/H] and their corresponding uncertainties from ASPCAP. These stars are Group A in Table \ref{AllStars} and the spectra for these objects are shown in Figure \ref{GroupASpec}. The discussion below is focused on these stars only.

\end{itemize}

\begin{table*}
\centering
\begin{adjustbox}{max width=\textwidth}
\begin{tabular}{ccccccccccccc}
\hline
ID & RA & Dec & \textit{V} & \textit{H} & Visits & $v_r$ & $\sigma_{v_r}$ & $T_{\rm eff}$ &
$\log g$ & [Fe/H] & [$\alpha$/Fe] & [C/Fe] \\
& (deg) & (deg) & (mag) & (mag) & & (km s$^{-1}$) & (km s$^{-1}$) & (K) & (dex) & (dex) & (dex) & (dex) \\
\hline
\textit{Group A}:\\
2M15312547+4220551  &  232.856165  &  42.348644  &  15.23  &  12.71  &  3  &  -152.6  &  1.7  &  4068  &  -0.15  &  -2.08  &  -0.28  &  0.75  \\ 
2M00114258+0109386  &  2.927455  &  1.160739  &  14.3  &  11.76  &  3  &  -154.1  &  0.1  &  4165  &  -0.14  &  -2.18  &  0.06  &  0.84  \\ 
2M21330683+1209406  &  323.278461  &  12.161281  &  -----  &  12.8  &  33  &  -294.2  &  0.4  &  4220  &  0.39  &  -2.01  &  0.29  &  0.91  \\ 
2M18111704-2352577  &  272.821009  &  -23.882698  &  -----  &  10.46  &  1  &  -129.1  &  -----  &  4238  &  0.65  &  -1.82  &  0.27  &  0.76  \\ 
2M16300629-1252459  &  247.526226  &  -12.879425  &  15.67  &  12.36  &  6  &  -296.2  &  0.6  &  4539  &  0.99  &  -1.95  &  0.30  &  0.97  \\ 
2M02121851+4923143  &  33.07714  &  49.387321  &  13.01  &  10.11  &  4  &  -141.5  &  0.1  &  4593  &  1.61  &  -1.82  &  0.44  &  0.82  \\
\hline
\textit{Group B}:\\
2M16334467-1343201  &  248.43614  &  -13.72225  &  13.8  &  10.18  &  3  &  19.4  &  0.4  &  4288  &  0.23  &  -2.1  &  0.2  &  0.89  \\ 
2M16562103+1002085  &  254.087657  &  10.035707  &  12.61  &  9.71  &  4  &  -149.7  &  0.3  &  4376  &  0.93  &  -1.98  &  0.38  &  0.88  \\ 
2M11584435+5518120  &  179.684833  &  55.303352  &  14.41  &  12.03  &  4  &  51.9  &  0.2  &  4555  &  1.29  &  -1.8  &  0.28  &  0.74  \\ 
2M16385680+3635073  &  249.736697  &  36.585381  &  15.35  &  12.98  &  13  &  -317.3  &  0.7  &  4561  &  1.46  &  -1.84  &  0.1  &  0.88  \\ 
2M14571988+1751501  &  224.332847  &  17.863941  &  13.91  &  11.32  &  3  &  -71.8  &  0.1  &  4566  &  1.35  &  -1.9  &  0.19  &  0.78  \\ 					
\hline
\textit{Group C}:\\
2M12473823-0814340  &  191.909322  &  -8.242799  &  15.66  &  13.56  &  13  &  -299.9  &  10.0  &  4317  &  0.79  &  -2.24  &  0.67  &  0.75  \\ 
2M05352696-0510173  &  83.862361  &  -5.171481  &  -----  &  10.93  &  4  &  -184.1  &  1.5  &  4591  &  2.34  &  -1.9  &  0.37  &  0.75  \\ 
\hline
\end{tabular}
\end{adjustbox}
\caption{List of APOGEE stars with [Fe/H]$<-1.8$, [C/Fe]$>+0.7$, and $T_{\rm eff} < 4600$K organized by the groups outlined in Section 3.1 and ordered by descending $T_{\rm eff}$. $V$-band magnitudes adopted from the \citet{Zacharias05} NOMAD Catalog.}
\label{AllStars}
\end{table*}

A search through the literature revealed that none of the Group A and B CEMP candidates have been previously identified. As a further test, we searched the complete APOGEE DR12 database for previously known CEMP stars. Stars with 2MASS identifiers in the data sampled from \citet{Lucatello05} (16 of 19 stars), \citet{Aoki07} (11 of 26 stars), \citet{Placco10} (11 of 18 stars),\citet{Hansen16no} (17 of 24) and \citet{Hansen16s} (19 of 22 stars) were queried in the APOGEE sample, however none of those 74 objects were found in DR12.

\section{ASPCAP elemental abundances}\label{sASPCAP}

\subsection{Carbon and nitrogen}\label{sCN}

The C and N abundances for the six Group A stars are presented in Table \ref{CNOTable}. Typical unmixed RGB halo stars with [Fe/H] $\sim -2.0$ show scaled-solar [C/Fe] and [N/Fe] ratios \citep{Gratton00, Cayrel04, Aoki07}. However, all of our Group A stars are C-enhanced and they have super-solar [N/Fe]. The N-enhancements in our stars are consistent with the abundance profiles observed in CEMP stars by \citet{Norris13}, \citet{Hansen15a}, and \citet{Hansen16a}. Of course, N enhancement in RGB stars is not unique to CEMP stars since mixing on the upper RGB can reduce the surface $^{12}$C abundance by a factor of $\sim2.5$, decrease $^{12}$C/$^{13}$C to less than 10, and increase [N/Fe] up to $\sim +0.4$ dex \citep{Gratton00, Cayrel04, Spite06, Aoki07, Placco14}. Surface Li can additionally act as a tracer for mixing, however the Li abundance cannot be determined from the APOGEE spectra.

\begin{figure}
\centering
\includegraphics[width=\linewidth]{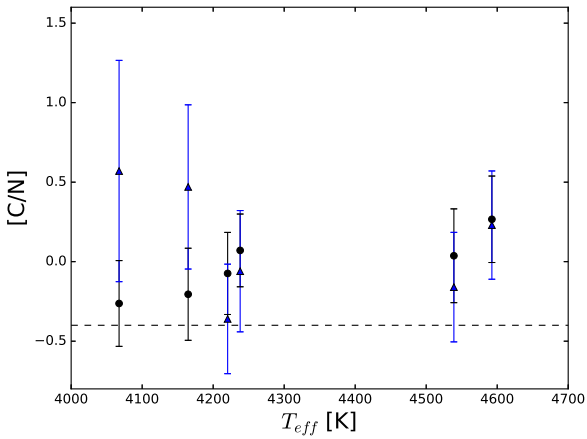}
\caption{[C/N] vs. $T_{\rm eff}$ for the Group A stars. Black circles correspond to the ASPCAP abundances and blue triangles correspond to the abundances derived in this work (see sections \ref{sMOOGErr}, \ref{sMOOGCNO}, and \ref{sCEMPConfirmed}). The dashed line at [C/N]$ = -0.4$ separates mixed ([C/N] $< -0.4$) and unmixed stars ([C/N] $> -0.4$) as prescribed by \citet{Spite05} for extremely metal-poor giants. The large error bars on the [C/N] abundance the two coolest stars derived in this work represent the combined 2$\sigma$ error as a result of difficulties in continuum placement. Despite the high likelihood of mixing in these stars considering they are on the RGB or AGB, the expression of unmixed abundance ratios is in favour of enhanced natal abundances in carbon.}

\label{CNMixing2}
\end{figure}

Addressing the possibility of mixing in our sample, Figure \ref{CNMixing2} shows [C/N] as a function of $T_{\rm eff}$ for our Group A stars. \citet{Spite05} examined the effects of mixing on the evolution of C and N in EMP giants and prescribed a separation between mixed and unmixed stars at [C/N] $= -0.4$.  All six Group A candidates lie above this cut at [C/N]$= -0.4$, suggesting the Group A stars are unmixed. However, this implication is unlikely considering all APOGEE targets are on the RGB or AGB where mixing processes and dredge-up are expected. The observation of unmixed abundance ratios, despite the high likelihood of mixing, is a sensible indicator that the natal carbon abundance of these stars was enhanced.

\begin{table*}
\caption{C and N Abundances for Group A stars as in the ASPCAP database}	
\begin{tabular}{lcccccc}
\hline
 &	2M15312547 & 2M00114258 & 2M21330683 & 2M18111704 & 2M16300629 & 2M02121851 \\
 & +4220551 & +0109386 & +1209406 & -2352577 & -1252459 & +4923143 \\
\hline
$T_{\rm eff}$ (K)		&		4068 $\pm$  91 	&		4165 $\pm$ 91 		&		4220  $\pm$  91		&		4238  $\pm$  91		&		4539  $\pm$  91		&		4593  $\pm$  91		\\
$\log g$		&		-0.15  $\pm$  0.11		&		-0.14  $\pm$ 0.11		&		0.39  $\pm$  0.11		&		0.65  $\pm$  0.11		&		0.99  $\pm$  0.11		&		1.61  $\pm$  0.11		\\
\text{[Fe/H]}		&		-2.08  $\pm$  0.06		&		-2.18  $\pm$  0.06		&		-2.01  $\pm$  0.06		&		-1.82  $\pm$  0.06		&		-1.95  $\pm$  0.06		&		-1.82  $\pm$  0.06		\\
\text{A(C)}		&		7.18  $\pm$  0.17		&		7.16  $\pm$  0.19		&		7.4  $\pm$  0.17		&		7.44  $\pm$  0.15		&		7.52  $\pm$  0.21		&		7.5  $\pm$  0.19		\\
\text{[C/Fe]}		&		0.75  $\pm$  0.17		&		0.84  $\pm$  0.19		&		0.91  $\pm$  0.17		&		0.76  $\pm$  0.15		&		0.97  $\pm$  0.21		&		0.82  $\pm$  0.19		\\
\text{A(N)}		&		6.8  $\pm$  0.23		&		6.73  $\pm$  0.24		&		6.83  $\pm$  0.22		&		6.73  $\pm$  0.21		&		6.85  $\pm$  0.24		&		6.59  $\pm$  0.22		\\
\text{[N/Fe]}		&		1.02  $\pm$  0.21		&		1.05  $\pm$  0.22		&		0.98  $\pm$  0.19		&		0.68  $\pm$  0.18		&		0.93  $\pm$  0.21		&		0.56  $\pm$  0.19		\\
\text{[C+N/Fe]}		&	0.82  $\pm$  0.16	&	0.89  $\pm$  0.17		&	0.92  $\pm$  0.16	&		0.74  $\pm$  0.15	&	0.96  $\pm$  0.19	&	0.78  $\pm$  0.19	\\
\hline
\end{tabular}
\label{CNOTable}
\end{table*}

Further supporting enhanced natal carbon abundances, the four warmer stars exhibit [C/Fe] $\leq$ [N/Fe] and [N/Fe] $> +0.4$ dex, which is unusual if CNO cycling was the only factor responsible for abundance variation. The two cooler stars, 2M15312547+4220551 and 2M00114258+0109386, have low [C/N] and [N/Fe] $>$ [C/Fe], which, coupled with their low surface temperatures and surface gravities, strongly indicate that they are evolved stars whose current abundances reflect CNO cycling on the upper RGB and potentially the third dredge-up as AGB stars. Even so, if these stars are expected to have experienced the highest degree of mixing, yet [C/N] $\geq -0.4$, their natal carbon abundance must be enhanced.

Another diagnostic of mixing vs. natal abundances is the sum of C+N+O.\footnote{We exclude O in our discussion as O is much less affected by CNO cycling on the RGB \citep{Yong08, Lardo12}.} [(C+N)/Fe] is expected to remain constant during a stars' evolution on the giant branch. At [Fe/H]$ = -2.0$, giant stars show $\langle$[(C+N)/Fe]$\rangle \sim 0$ \citep{Spite06, Lardo15}. As metallicity decreases, the average [(C+N)/Fe] in the giants is seen to vary from solar to higher values, e.g. at [Fe/H] $\leq -3.0$,  $\langle[$(C+N)/Fe]$\rangle \sim +0.5$ For all stars in our group Group A sample, [(C+N)/Fe]$ > +0.7$ with a group mean at  $\langle$[(C+N)/Fe]$\rangle =  +0.85 \pm 0.10$. This is much higher than expected for typical metal-poor field stars at these metallicities. In summary, the high mean [(C+N)/Fe] supports enriched natal abundances in line with other CEMP stars.

Carbon abundances may be enriched in intermediate-mass metal-poor AGB stars as a result of the third dredge-up, mimicking CEMP-$s$ abundance profiles \citep{Herwig05, Cristallo11, Bisterzo12, Hansen16a}. To eliminate the possibility that our Group A stars are highly self-enriched AGB stars, their luminosities, neutron-capture element abundances, and $^{12}C/^{13}C$ are needed. Unfortunately strong $^{13}C$ molecular features, as well as neutron-capture lines are inaccessible in the APOGEE spectra, warranting the need for follow-up observations in the optical regime.

\subsection{$\alpha$-elements}\label{salpha}

In ASPCAP, [$\alpha$/M] serves as a free parameter in the synthetic spectral grid when performing its least squares minimization process. From the global [$\alpha$/M], the abundances of the $\alpha$-elements O, Mg, Si, S, Ca, and Ti are derived a posteriori. Keeping all other parameters fixed, the [$\alpha$/M] dimension of the spectral grid used by ASPCAP is varied, and, by examining weighted spectral "windows" (regions in the spectrum sensitive to a particular species) an estimate for the individual abundance is found \citep{GarciaPerez15, Holtzman15}. As discussed in Section \ref{sSample}, we adopt the Fe based metallicities therefore we have calculated [$\alpha$/Fe] from the [Fe/H]-[M/H] offset per star (see Table \ref{AlphaTable}). Ti has been removed from this analysis as a result of observed deviations from the expected trend in metallicity within the APOGEE data \citep{Holtzman15, Hawkins16}. 

\begin{table*}
\caption{$\alpha$-abundances for Group A stars as in the ASPCAP database. $\sigma$ is from the weighted average of all $\alpha$-elements here with the exception of S in 2M15312547+4220551 and Ca in 2M00114258+0109386. An empty entry means no abundance was determined by ASPCAP.}
\begin{tabular}{ccccccc}
\hline
 &	2M15312547 & 2M00114258 & 2M21330683 & 2M18111704 & 2M16300629 & 2M02121851 \\
 & +4220551 & +0109386 & +1209406 & -2352577 & -1252459 & +4923143 \\
\hline
\text{[O/Fe]}	&	-0.15  $\pm$  0.09	&	-0.01  $\pm$  0.10	&	0.29  $\pm$  0.09	&	0.28  $\pm$  0.08	&	0.32  $\pm$  0.10	&	0.45  $\pm$  0.09	\\
\text{[Mg/Fe]}	&	-0.08  $\pm$  0.16	&	0.19  $\pm$  0.16	&	0.17  $\pm$  0.12	&	0.22  $\pm$  0.12	&	0.05  $\pm$  0.14	&	0.21  $\pm$  0.12	\\
\text{[Si/Fe]}	&	-0.39  $\pm$  0.09	&	0.05  $\pm$  0.09	&	0.27  $\pm$  0.08	&	0.26  $\pm$  0.08	&	0.30  $\pm$  0.09	&	0.47  $\pm$  0.08	\\
\text{[S/Fe]}	&	0.69  $\pm$  0.16	&	0.29  $\pm$  0.16	&	0.36  $\pm$  0.12	&	0.37  $\pm$  0.12	&	0.49  $\pm$  0.16	&	0.63  $\pm$  0.13	\\
\text{[Ca/Fe]}	&	-0.67  $\pm$  0.24	&	-----	&	0.50  $\pm$  0.15	&	0.23  $\pm$  0.15	&	0.37  $\pm$  0.20	&	0.35  $\pm$  0.14	\\
\text{[$\alpha$/Fe]}	&	-0.28  $\pm$  0.13	&	0.06  $\pm$  0.09  	&	0.29  $\pm$  0.08  	&	0.27  $\pm$  0.06  	&	0.3  $\pm$  0.09  	&	0.44  $\pm$  0.09\\ 
\hline
\end{tabular}
\label{AlphaTable}
\end{table*}

\begin{figure}
\includegraphics[width=\linewidth]{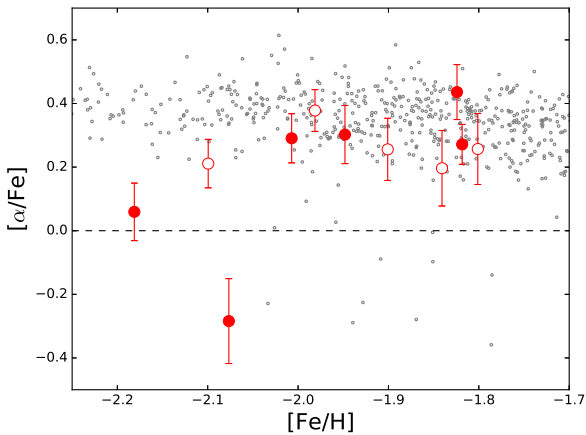}
\caption{[$\alpha$/Fe] vs. [Fe/H] for the Group A and Group B CEMP candidates. [$\alpha$/Fe] is calculated as the uncertainty weighted average of the calibrated ASPCAP abundances for Mg, Si, S, and Ca (with the exception of S in 2M15312547+4220551 and Ca in 2M00114258+0109386.) Grey dots correspond to typical metal-poor stars from APOGEE and normal field and halo stars \citep{Venn04, Frebel10}, solid red circles correspond to our Group A candidates and open red circles to Group B. The dashed line at [$\alpha$/Fe] = 0.0 separates the $\alpha$-poor stars from the bulk of the sample.}
\label{AlphaPlot}
\end{figure}

Examining the CEMP candidates against background field stars, we compare [$\alpha$/Fe] for Groups A and B to the $\alpha$ abundances of bulk APOGEE field stars and to metal-poor stars in the Galactic halo and disk summarized by \citet{Venn04} and \citet{Frebel10} in Figure \ref{AlphaPlot}. The weighted average for the Group A stars is [$\alpha$/Fe] = $+0.23 \pm 0.23$, consistent with the expected enhancement of [$\alpha$/Fe]$\sim +0.30$ dex for stars in the halo, indicating the majority of our sample are typical Galactic halo stars.

Two stars, 2M15312547+4220551 and 2M00114258+0109386, display $\alpha$-abundances substantially lower  than typical values found for halo stars at [Fe/H]$\sim -2.0$ ([$\alpha$/Fe] = $-0.28 \pm 0.13$ and $0.06 \pm 0.09$, respectively). The individual $\alpha$-abundances for these objects can be examined in Table \ref{AlphaTable}. The abundances of O, Mg, Si, and Ca for 2M15312547+4220551 are all sub-solar with the exception of [S/Fe] = $+0.69 \pm 0.16 $, which we expect this is the result of an unrecognised blend (see Section \ref{sFERRELimits}) and have ignored it here. The S abundance of 2M00114258+0109386 is also affected by nearby blends and a Ca abundance could not be determined by ASPCAP, however the abundances of O, Mg, and Si suggest this star is $\alpha$-challenged. The discovery of $\alpha$-challenged stars within the Galaxy may serve as a link to the remnants of accreted dwarf galaxies in the context of Galactic evolution (see Section \ref{sdSph} for further discussion.)

\subsection{ASPCAP/FERRE Limitations}\label{sFERRELimits}

Using the APOGEE-specific implementation of FERRE from \citet{Bovy16a}\footnote{This package is available at https://github.com/jobovy/apogee} which includes the renormalized synthetic spectra and model atmospheres used by ASPCAP, we examine the abundances of individual spectral features and their ASPCAP weighting kernels. We also investigate nearby line contamination, upper limits, and unreported data quality issues. FERRE and ASPCAP give each spectral line a relative weighting kernel determined through a combination of methods including comparison to the solar spectrum, to the Arcturus spectrum, and by line fitting across the complete APOGEE sample \citep{Shetrone15}. 

\begin{figure*}
\includegraphics[width=\linewidth]{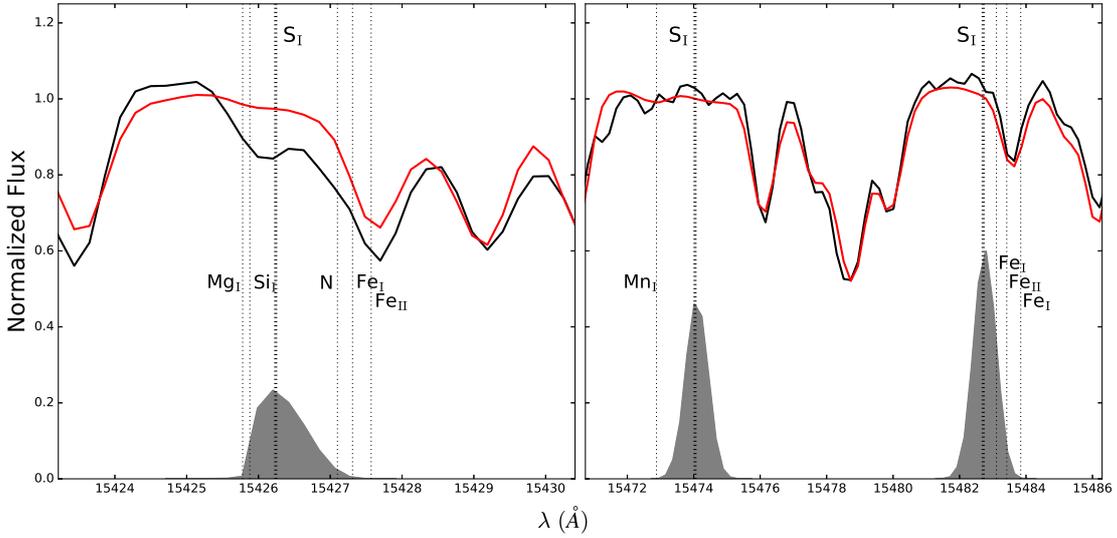}
\caption{FERRE windows of the S features used to estimate a S abundance for the $\alpha$-poor star 2M15312547+4220551 ([Fe/H] = $-2.08 \pm 0.06$, [$\alpha$/Fe] = $-0.28 \pm 0.13$, [S/Fe] = $+0.69 \pm 0.16$). The APOGEE combined spectrum is in black and the ASPCAP synthetic spectrum in red. The weighing kernels used by ASPCAP/FERRE to assign a relative weight to each spectral feature when determining an overall abundance are shown in filled grey. The kernel width often results in nearby lines contributing to the perceived strength of the line of interest, increasing the estimated abundance. The width of the weighting kernel allows nearby metal lines to add to the estimated S abundance. The lack of a S line at 15,474 \AA\ is in favour of a lower than reported S abundance.}
\label{FERRE_S}
\end{figure*}

For nearly every star in Group A, we found that the weighting scheme was not ideal, based on the spectral quality and other poor synthetic fits. As an example, the spectral windows used by ASPCAP/FERRE to estimate the S abundance for 2M15312547+4220551 are shown in Figure \ref{FERRE_S}. Based on the width of the \ion{S}{i} weighting kernel, nearby lines of Fe and OH are captured by the S weighting kernel. This results in a lower apparent flux in that spectral region than if the S line was the only dominant species which, in turn, forces a stronger line depth in the synthetic spectral fit. The lack of a \ion{S}{i} line at 15,746 \AA\ further supports a lower S abundance.

The challenge of analysing metal-poor and chemically peculiar stars with ASPCAP is not unexpected. ASPCAP covers a large multi-dimensional parameter space which is unavoidably subject to edge effects. CEMP stars exist not only at this edge, but in a parameter space that is still weakly constrained by existing data. The ASPCAP weighting scheme is a compromise to fit the features of stars over a wide range of stellar parameters, but it does not appear to apply to the most metal-poor stars, even those with the highest data quality. Consequently, we suspect all the elemental abundances in the DR12 ASPCAP database for the carbon-enhanced stars with [Fe/H]$\lesssim -2.0$ are poorly determined.

In the next section, we explore an independent model atmospheres analysis of our Group A CEMP candidates.

\section{MOOG Spectrum Synthesis}\label{sMOOG}

For the six Group A CEMP candidates, we carry out a detailed model atmospheres analysis of the APOGEE spectra using the \textsc{moog} radiative transfer code\footnote{\textsc{moog} was originally written by Chris Sneden (1973), and has been updated and maintained, with the current versions available at http://www.as.utexas.edu/$\sim$chris/moog.html.}. The APOGEE DR12 linelist \citep{Shetrone15} was used for the LTE radiative transfer calculations. Spherically-symmetric, moderately CN-cycled and $\alpha$-enriched ([C/Fe]$ = -0.13$, [N/Fe]$ = +0.31$, [O/Fe]$ = +0.40$) model atmospheres were adopted from the MARCS grid \citep{MARCS}. Solar abundance ratios were adopted from \citet{Asplund09} however solar C, N, and O were taken from \citet{Caffau11}. As with the ATLAS9 models used by APOGEE, the MARCS models have limited coverage for carbon-enhanced atmospheres and thus the carbon abundances were increased in \textsc{moog} to resemble a CEMP atmosphere. 

The calibrated \texttt{aspcapStar} stellar parameters $T_{\rm eff},\ \log g,$ microturbulence, and [Fe/H] are required for a model atmospheres analysis and have been adopted directly from the ASPCAP pipeline. Furthermore, we do not have distances to these stars in order to use the color-temperature-metallicity calibrations, e.g. \citet{Casagrande10} to independently derive stellar parameters. 

\begin{figure}
\includegraphics[width=\linewidth]{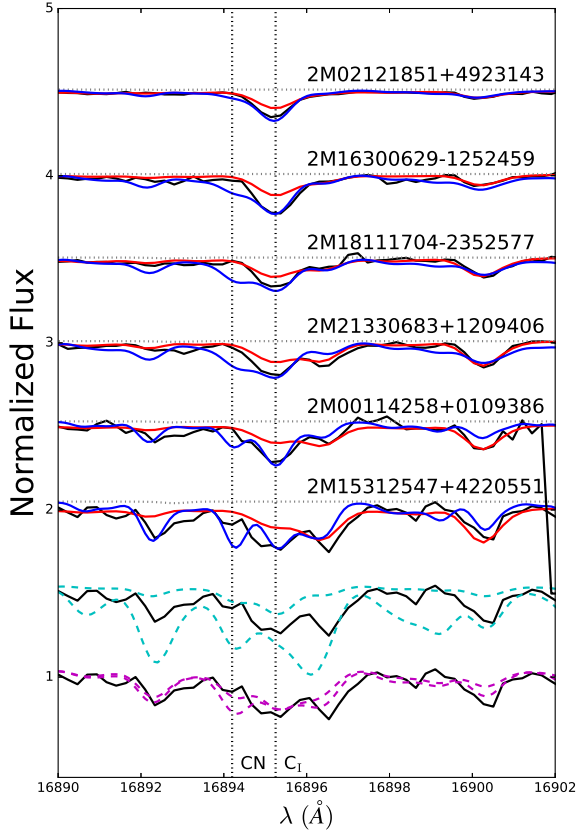}
\caption{Spectra of Group A stars centred around the atomic \ion{C}{i} line. The APOGEE combined spectrum (black) with the ASPCAP synthetic spectrum (red) is shown in comparison to the best fit synthetic spectra derived using \textsc{moog} based on the ASPCAP stellar parameters (blue). Offsets in flux are arbitrary. The bottom two spectra represent the $\pm 1\sigma$ variations in [C/Fe] (dashed cyan) and [N/Fe] (dashed magenta) for 2M15312547+4220551. Variations in [C/Fe] and [N/Fe] are comparable for the other stars in our sample. Table \ref{ASPCAPvMOOG} summarizes the measured abundances.}
\label{SynthSpecASPCAP}
\end{figure}

The results of our synthesis for all Group A stars are shown in Figure \ref{SynthSpecASPCAP}, centred around the atomic \ion{C}{i} line at 16895 \AA\ and the derived abundances versus those returned by ASPCAP for all Group A stars are compared in Table \ref{ASPCAPvMOOG}. The following sections discuss the measured abundances for Fe, C, N, O, and the $\alpha$-elements Mg, Si, S, and Ca and their comparisons to the ASPCAP results.

\subsection{Abundance uncertainties}\label{sMOOGErr}

The abundance uncertainties in this work reflect both the systematic errors introduced through uncertainties in the adopted stellar parameters, as well as the random intrinsic errors associated with line-to-line abundance variations. Typical reported uncertainties in the ASPCAP stellar parameters $T_{\rm eff},\ \log g,$ microturbulent velocity $\xi$, and [Fe/H] are on the order of $\pm 100$K, $\pm 0.1$dex, $\pm  0.2$km/s
\footnote{\citet{Holtzman15} shows an rms scatter in $\xi$ vs. $\log g$ of 0.22 km s$^{-1}$ for stars with $\log g < 3.8$.}, and $\pm 0.1$dex, respectively. The effect of these model atmosphere uncertainties on the derived abundances is shown in Table \ref{SysErrs} with the total systematic uncertainty determined by adding the sensitivities from each stellar parameter in quadrature (ignoring covariances). The effect of these uncertainties was investigated for two stars in our sample, 2M15312547+4220551 and 2M02121851+4923143, to span our full temperature range. Systematic errors are  higher for the cooler star by $\sim 0.1$dex which is reasonable considering the stronger effects of molecular blends and subsequent difficulty in continuum placement at lower temperature. Owing to similar temperatures and $SNR$, the systematic abundance uncertainties found for 2M15312547+4220551 ($T_{\rm eff} = 4068$K) were also applied to 2M00114258+0109386 ($T_{\rm eff} = 4165$K). The systematic uncertainties for 2M02121851+4923143 ($T_{\rm eff} = 4593$K) were applied to the remaining three stars in our sample.

\begin{table}
\caption{Systematic Abundance Sensitivities. Errors shown here were added in quadrature to calculate the total systematic uncertainty and do not reflect the uncertainty introduced by line-to-line scatter. See Table \ref{ASPCAPvMOOG} for line-to-line measurement errors. Since S and Ca abundances could not be determined for 2M15312547+4220551, systematic errors were not investigated.}
\begin{tabular}{lrrrrr}

\hline
 & $\Delta T_{\rm eff}$ &  $\Delta \log$g & $\Delta \xi $ & $\Delta$[Fe/H]& Total \\
 & $+$100 K & $+0.1$ dex & $+$0.5 km s$^{-1}$ & $+0.1$ dex & \\
\hline 

\multicolumn{5}{l}{2M15312547+4220551 ($T_{\rm eff} = 4068$ K) } 		&		\\
\text{[C/Fe]}	&	0.20	&	-0.05	&	-0.04	&	-0.04	&	0.21	\\
\text{[N/Fe]}	&	-0.14	&	-0.03	&	-0.02	&	-0.04	&	0.15	\\
\text{[O/Fe]}	&	-0.16	&	-0.11	&	-0.02	&	0.10	&	0.22	\\
\text{[Mg/Fe]}	&	-0.25	&	-0.11	&	-0.06	&	-0.08	&	0.29	\\
\text{[Si/Fe]}	&	-0.20	&	0.15	&	0.08	&	0.08	&	0.27	\\
\\											
\multicolumn{5}{l}{2M02121851+4923143 ($T_{\rm eff} = 4593$ K) } 		&		\\
\text{[C/Fe]}	&	0.20	&	-0.05	&	-0.04	&	-0.03	&	0.21	\\
\text{[N/Fe]}	&	0.17	&	-0.06	&	0.02	&	-0.02	&	0.18	\\
\text{[O/Fe]}	&	-0.13	&	-0.10	&	-0.01	&	-0.02	&	0.16	\\
\text{[Mg/Fe]}	&	0.05	&	-0.07	&	-0.02	&	0.02	&	0.09	\\
\text{[Si/Fe]}	&	0.08	&	0.09	&	0.05	&	-0.04	&	0.14	\\
\text{[S/Fe]}	&	-0.20	&	-0.05	&	-0.04	&	-0.04	&	0.21	\\
\text{[Ca/Fe]}	&	-0.10	&	-0.10	&	-0.08	&	-0.06	&	0.17	\\
\hline
\end{tabular}
\label{SysErrs}
\end{table}
 
Intrinsic abundance uncertainties were determined for each spectral line by first calculating the \textit{observed - synthetic} residual over a spectral feature, and by then adjusting the abundance until this residual was comparable in magnitude to deviations in the continuum noise.  The total intrinsic measurement error for a species $X$ was then determined as the mean in the line-to-line scatter $\sigma_X / \sqrt{N_X}$. These abundance uncertainties are reflected in Table \ref{ASPCAPvMOOG}. The systematic uncertainties and intrinsic uncertainties are added in quadrature to calculate the total abundance uncertainty used in the discussion and Figures \ref{CNMixing2} and \ref{SynthSpecASPCAP}.

\subsection{Fe abundances}\label{sMOOGFe}

Initially, we examined the ASPCAP synthetic spectra and found the general fits to the Fe lines to quite good, and therefore we expect the FERRE stellar parameters ($T_{\rm eff}$, $\log g$, and [Fe/H]) to be in fair agreement. As an iteration on this, we also examined isolated \ion{Fe}{i} lines in each Group A star by-eye. Averaging over the 54 \ion{Fe}{i} lines used by ASPCAP to determine the abundance of Fe, we still did not observe any sign of deviation in the spectra or the synthetic fits. We adopt [Fe/H] from ASPCAP as an initial stellar parameter in our synthetic synthesis.

Testing this further, the isolated \ion{Fe}{i} features were synthesized in \textsc{moog} for a sample star, 2M02121851+4923143. This star was selected as its temperature is the highest in our sample ($T_{\rm eff} = 4593$K) and we would expect the degeneracy between $T_{\rm eff}$ and [Fe/H] to be the most prominent in this spectrum as high $T_{\rm eff}$ and low metallicity produce similar observed effects in the spectra (i.e. weak Fe lines). In Table \ref{FeLinelist}, individual \ion{Fe}{i}  abundances are shown and the mean Fe abundance was determined through a weighted average of all the individual lines, with the weights given by $1/ \sigma^2_i$ where $\sigma_i$ is the abundance uncertainty on each line (see Section \ref{sMOOGErr}). The metallicity from \textsc{moog} was determined as [Fe/H] = $-1.7 \pm 0.1$ based on 21 FeI lines. ASPCAP reports [Fe/H] = $-1.82 \pm 0.06$ for this star, consistent within the 1$\sigma$ measurement errors to the results of this work.

\begin{table}
\centering
\caption{\ion{Fe}{i} lines measured in 2M02121851+4923143}
\begin{tabular}{cccc}
\hline
Wavelength (\AA) & $\chi$(eV) & $\log \ gf$ & Log abundance \\
\hline
15211.75	&	6.323	&	-2.285	&	5.83	\\
15223.78	&	5.587	&	-0.082	&	5.93	\\
15298.74	&	5.308	&	0.65	&	5.88	\\
15595.76	&	6.364	&	0.659	&	5.83	\\
15625.92	&	5.539	&	0.253	&	5.78	\\
15636.22	&	5.351	&	-0.114	&	5.93	\\
15681.81	&	6.246	&	0.157	&	5.98	\\
15727.88	&	5.62	&	-0.03	&	5.98	\\
15899.57	&	6.258	&	0.222	&	5.68	\\
15969.23	&	5.921	&	-0.128	&	5.48	\\
16045.04	&	5.874	&	0.066	&	5.68	\\
16106.81	&	5.874	&	0.196	&	5.88	\\
16130.31	&	6.351	&	0.624	&	5.78	\\
16157.66	&	5.351	&	-0.743	&	5.68	\\
16169.45	&	6.319	&	0.723	&	5.83	\\
16320.78	&	6.28	&	0.871	&	5.88	\\
16521.74	&	6.286	&	0.5	&	5.78	\\
16528.98	&	6.336	&	0.577	&	5.63	\\
16566.29	&	5.978	&	0.065	&	5.78	\\
\hline
\label{FeLinelist}
\end{tabular}
\end{table}

\subsection{Carbon, Nitrogen, and Oxygen}\label{sMOOGCNO}

ASPCAP determines C and N abundances by simultaneously fitting multiple CN and CO molecular features over the full spectral range. Oxygen is later determined from the $\alpha$-abundance as discussed in Section \ref{salpha} and \ref{sMOOGalpha}. Since multiple metals contribute to the strength and structure of these molecular bands, the method employed by ASPCAP to determine the abundances of C, N, and O can not properly handle the degeneracies that may exist in the abundances of these three species. 

To break these possible degeneracies, we follow the process established by \citet{Smith13} and \citet{Lamb15} and use a line-by-line procedure to derive the C, N, and O abundances. Isolated atomic and molecular features were synthesized and the model atmosphere was updated to incorporate the new abundance estimate. Using the atomic \ion{C}{i} line at 16895 \AA\, an initial carbon abundance was found and used as a seed for an O abundance estimate. OH lines with no CN contamination were then synthesized to measure the oxygen abundance. Adopting mean O abundance, C was re-determined from the \ion{C}{i} line alone. This process was iterated until the C and O abundances converged, typically requiring two iterations. Finally, using the best C estimate, N was measured via CN lines between 15100 and 16100 \AA. The complete line list is given in Table \ref{CNOLinelist}. The mean abundance and final error for each species was calculated in similar fashion to the iron lines in Section \ref{sMOOGFe}.

We chose to focus only on the atomic \ion{C}{i} line for a few reasons.  At high C, N, and O abundances, like those expected in the CEMP stars, the structure of the CO and CN molecular features is highly degenerate, prohibitive in determining abundances with any reasonable uncertainty. Temperature effects magnify these issues as well: at low $T_{\rm eff}$ ($\lesssim 4000$K) these molecular features dominate the spectra resulting in strong molecular blends, making it exceptionally difficult to separate the effects each species has on the spectral features as well as complicating the placement of the continuum. In stars with temperatures greater than $\sim 4600$K, these molecular features disappear as discussed in Section \ref{sUncert}. The atomic line is solely dependent on the C abundance and thus the effects of N, O, and $T_{\rm eff}$ seen in the molecular bands is not observed. \footnote{The atomic data available for the \ion{C}{i} transition is high fidelity and has been assigned a 90\% confidence rating by NIST. Manually changing $\log gf$ in our linelist by 10\%, to account for the small uncertainty in atomic data, results in variations in the derived C abundance that are imperceivable ($\lesssim 0.05$ dex) considering the quality of our spectra.}
High $T_{\rm eff} >$ 4600 K remains an issue, however these stars have been removed from our sample as discussed in sections \ref{sUncert} and \ref{sSample}.

Further supporting our choice of using the atomic line alone, in their analysis of four metal-poor giant stars in globular clusters, \citet{Lamb15} incorporated the atomic \ion{C}{i} line into their determination of [C/Fe]. Comparing the C abundance derived from the atomic line alone (see their supplementary atomic and molecular line list) to their global, NLTE corrected C abundance for each star (their Table 7, adjusted for a -0.1 dex NLTE offset in [Fe/H]), the differences are on the order of 0.1 dex or less. While this is a small sample, these results suggest the atomic line alone may provide an accurate estimation of the global C abundance when the degeneracy between C, N, and O is properly handled.

Finally, the atomic \ion{C}{i} line is not used by APOGEE when estimating a C abundance, presumably due to the typically lower $SNR$ at this wavelength in the APOGEE spectra. While the $SNR$ around this line in our candidate spectra provides the source for our large uncertainties in our reported C abundance, we are able to clearly see this line in our Group A spectra, allowing us to determine a C abundance through a method independent of APOGEE. 

The carbon abundances derived through \textsc{moog} were found to be within the uncertainties of the C-abundance reported by ASPCAP for the four warmer stars in our sample. Typical deviations in [C/Fe]$_{This Work} -$[C/Fe]$_{ASPCAP}$ for these four stars were on the order of 0.1 dex, half of the standard error reported for this abundance. The two cooler stars show much larger deviations in [C/Fe] which we attribute to the difficulty in continuuum placement at these lower temperatures and thus we chose to adopt the 2$\sigma$ errors for abundances of these stars. The agreement between our work and the ASPCAP results cautiously supports the reliability of the reported abundance for our Group A stars.

Oxygen abundances were determined using ~25 OH lines per star, spanning the full range of the APOGEE spectra. Again, owing to the degeneracies between C, N, and O, the molecular CO bands were avoided when possible while measuring O. The O abundances reported in this work are systematically higher than the ASPCAP abundances, with differences up to 0.5 dex. We suspect this discrepancy is a consequence of continuum placement and the simultaneous estimations of C, N, and O by ASPCAP, rather than following the iterative procedure of \citet{Smith13}. \citet{Lamb15} highlights the sensitivity of the abundances derived for these species when analysing infrared spectra.

Nitrogen abundances were determined through synthetic spectral fits of the CN electronic transition lines, primarily in the "blue" chip, after the C and O abundances were determined. The N abundances measured for the warmest Group A stars are consistent with the ASPCAP values to within the 1$\sigma$ uncertainties, however a systematic increase in our measured abundance is observed for the two coolest stars. At lower temperatures, the degeneracy between C and N in these CN lines becomes stronger. Despite a reportedly high $SNR$ and no visible persistence issues, the spectral quality over the "blue" chip for these two cooler stars is poor ($SNR \sim 20$) and the measured abundances may also be affected by issues in normalization and placement of the continuum. We additionally note the poor synthetic fit to the CN band at 16894 \AA\ in Figure \ref{SynthSpecASPCAP} for all Group A stars. This implies the reported N abundance is too high, however this is presumably due to poor atomic data for this particular transition. The small degree of line-to-line scatter in the N abundance derived by the multiple CN lines in the `blue` chip supports higher fidelity atomic data for those transitions and supports the reliability of the reported abundance.

\begin{table*}
\caption{Comparison of Group A abundances derived from ASPCAP and \textsc{moog} spectrum syntheses. Abundance errors shown here reflect the measurement error $\sigma / \sqrt{N}$ where $\sigma$ is the standard deviation of the line-to-line scatter and $N$ is the number of lines used in calculating an average abundance. Systematic errors as a result of uncertainties in the stellar parameters are shown in Table \ref{SysErrs}. A blank entry means no abundance could be reliably determined as a result of the quality of the spectra.}

\begin{tabular}{cccccccc}
\hline
Species & Source &	2M15312547 & 2M00114258 & 2M21330683 & 2M18111704 & 2M16300629 & 2M02121851 \\
 & & +4220551 & +0109386 & +1209406 & -2352577 & -1252459 & +4923143 \\
\hline

\text{[Fe/H]}	&	ASPCAP	&	
-2.08  $\pm$  0.06	&	-2.18  $\pm$  0.06	&	-2.01  $\pm$  0.06	&	-1.82  $\pm$  0.06	&	-1.95  $\pm$  0.06	&	-1.82  $\pm$  0.06	\\
\text{[C/Fe]}	&	ASPCAP	&	
0.75  $\pm$  0.17	&	0.84  $\pm$  0.19	&	0.91  $\pm$  0.17	&	0.76  $\pm$  0.15	&	0.97  $\pm$  0.21	&	0.82  $\pm$  0.19	\\
	&	This Work	& 1.25 $\pm$ 0.30	&	1.06 $\pm$ 0.20	&	0.98  $\pm$  0.15	&	0.77  $\pm$  0.20	&	1.11  $\pm$  0.15	&	0.75  $\pm$  0.13	\\
	
\text{[N/Fe]}	&	ASPCAP	&	
1.02  $\pm$  0.21	&	1.05  $\pm$  0.22	&	0.98  $\pm$  0.19	&	0.68  $\pm$  0.18	&	0.93  $\pm$  0.21	&	0.56  $\pm$  0.19	\\
	&	This Work	&	0.67 $\pm$ 0.12 	& 	0.59 $\pm$ 0.10	&	1.34  $\pm$  0.14	&	0.83  $\pm$  0.17	&	1.27  $\pm$  0.14	&	0.52  $\pm$  0.15	\\
	
\text{[O/Fe]}	&	ASPCAP	&	
-0.15  $\pm$  0.09	&	-0.01  $\pm$  0.10	&	0.29  $\pm$  0.09	&	0.28  $\pm$  0.08	&	0.32  $\pm$  0.10	&	0.45  $\pm$  0.09	\\
	&	This Work	&	0.15 $\pm$ 0.14 	&	0.24 $\pm$ 0.12	&	0.79  $\pm$  0.10	&	0.82  $\pm$  0.13	&	0.89  $\pm$  0.08	&	0.42  $\pm$  0.10	\\
	
\text{[Mg/Fe]}	&	ASPCAP	&	
-0.08  $\pm$  0.16	&	0.19  $\pm$  0.16	&	0.17  $\pm$  0.12	&	0.22  $\pm$  0.12	&	0.05  $\pm$  0.14	&	0.21  $\pm$  0.12	\\
	&	This Work	&	-0.57  $\pm$  0.22	&	-0.48  $\pm$  0.18	&	-0.19 $\pm$  0.14	&	-0.05 $\pm$ 0.15	&	-0.10 $\pm$ 0.15	&	0.00  $\pm$  0.13	\\
	
\text{[Si/Fe]}	&	ASPCAP	&	
-0.39  $\pm$  0.09	&	0.05  $\pm$  0.09	&	0.27  $\pm$  0.08	&	0.26  $\pm$  0.08	&	0.3  $\pm$  0.09	&	0.47  $\pm$  0.08	\\
	&	This Work	&	$< 0.33$	&	-0.39  $\pm$  0.24	&	0.14 $\pm$  0.17	&	0.12 $\pm$ 0.17	&	0.05 $\pm$ 0.14	&	0.08  $\pm$  0.13	\\
	
\text{[S/Fe]	} &	ASPCAP	&	
0.69  $\pm$  0.16	&	0.29  $\pm$  0.16	&	0.36  $\pm$  0.12	&	0.37  $\pm$  0.12	&	0.49  $\pm$  0.16	&	0.63  $\pm$  0.13	\\
	&	This Work	&	-----	&	-----	&	-----	&	-----	&	-----	&	$< 0.17$	\\
	
\text{[Ca/Fe]}	&	ASPCAP	&	
-0.67  $\pm$  0.24	&	-----	&	0.50  $\pm$  0.15	&	0.23  $\pm$  0.15	&	0.37  $\pm$  0.20	&	0.35  $\pm$  0.14	\\
	&	This Work	&	-----	&	-----	& ----- &	----- & ----- &	$< 0.07$ \\
\hline
\end{tabular}
\label{ASPCAPvMOOG}
\end{table*}

\subsection{$\alpha$-elements}\label{sMOOGalpha}
Abundances for the $\alpha$-elements Mg, Si, S, and Ca were determined through the \textsc{moog} synthetic spectrum analysis. Similar to O, much larger variations between the ASPCAP abundances and those determined in this work are observed than for C and N. The primary source for these discrepancies resides in the method ASPCAP uses to determine the abundances for the $\alpha$-elements. As discussed in Section \ref{salpha}, ASPCAP uses [$\alpha$/M] as a free parameter to generate a model spectrum and then scales the abundance of each $\alpha$-element based on the line-list and weighting windows for each element. This process works well if the abundance of a particular $\alpha$-element does not vary  from the overall $\alpha$-abundance and if the weighting kernels are dominated by the species of interest (however we have shown that even this can fail, e.g. the isolated \ion{S}{I} lines in 2M15312547+4220551.) By treating each line and each species individually, we arrive at a less biased abundance estimate. We are also able to determine upper limits in our process and identify when the ASPCAP abundance is not a determination, but likely an upper limit. The line list for these species is given in Table \ref{AlphaLinelist}.

\textit{Magnesium}: Mg abundances were determined using three lines between 15745 \AA\ and 15770 \AA. These lines were also used to set the Gaussian broadening parameter in our synthetic spectra and provided a clear estimate for the placement of the continuum. Comparing the ASPCAP abundances to our \textsc{moog} results shows that ours are typically lower by $\sim 0.3$ dex. We suggest this reflects issues in the continuum normalization of the ASPCAP spectra or under-represented systematic errors when adopting the ASPCAP stellar parameters (see Section \ref{sMOOGErr} for typical errors reported by ASPCAP).

\textit{Silicon}: Six lines spanning the "green" and "red" chips were used to determine the Si abundance for our Group A stars. Si was found to be lower than what was reported by APOGEE by $\sim 0.3$ dex and only an upper limit could be determined for 2M1531257+4220551. Large line-to-line scatter up to 0.5 dex was observed, particularly in the cooler stars where continuum placement became more spurious as a result of numerous molecular features.

\textit{Sulphur and Calcium}: The abundances of S and Ca could not be reliably determined for any of the Group A stars. Low $SNR$, weak spectral features, poor normalization of the combined spectra, and blending with nearby Fe lines were prohibitive in identifying the presence of a line. With the exception of 2M02121851+4923143, poor continuum normalization prevented even upper limit determinations for the Group A stars.

\section{Discussion}\label{sDiscussion}

\subsection{CEMP stars confirmed}\label{sCEMPConfirmed}

Through our synthetic spectrum analysis, carbon-enhancement has been verified for the Group A candidates to within the precision of the stellar parameters and model atmospheres used by APOGEE. Since the N abundances determined in this work have undergone a shift to higher [N/Fe] than reported by ASPCAP, and since the interpretation of the nature of CEMP stars is heavily reliant on understanding natal vs. evolved abundances as explored in Section \ref{sCN}, we re-examine the possibility of mixing within our sample. Figure \ref{CNMixing2} includes the [C/N] ratios derived in this work for the Group A stars. Following the discussion in Section \ref{sCN}, the four warmer objects in our sample appear to remain as unmixed giants, despite the high likelihood of mixing on the RGB. This strongly supports a high natal carbon abundance in these stars. The nature of the two cooler stars however is less clear. As suggested in Section \ref{sMOOGCNO}, molecular blends obfuscate the spectra at low temperature, making it exceedingly difficult to accurately place the continuum when comparing synthetic spectra to data. While the abundances of C, N, and [C/N] are still higher than expected for typical mixed metal-poor giants, the consequences of mixing are difficult to interpret until higher precision abundances are available.

We have additionally estimated the carbon corrections of our Group A stars following the procedure developed by \citet{Placco14}.  These corrections work to account for CN processing on the upper RGB and return a natal carbon abundance based on stellar parameters. Using their online calculator\footnote{http://www3.nd.edu/~vplacco/carbon-cor-input.html}, we find the carbon corrections for the Group A stars are typically $\delta$[C/Fe]$ = +0.2 $dex. While this further supports the assertion the Group A stars are truly carbon-enhanced, the corrections are within our combined errors on [C/Fe]. Since we are limited by systematics when attempting to determine more precise abundances, to remain consistent with the ASPCAP labels we choose to not adopt the carbon corrections in our current discussion. It is worth noting more CEMP stars are likely to be found within APOGEE if these corrections are applied homogeneously across the database. As an example, stars with $\log g = 1.0$, [Fe/H]$=-2.0$, and [C/Fe]$_{ASPCAP}=+0.5$ can be corrected up to a natal [C/Fe]$=+0.9$ using the carbon correction calculator which would classify them as CEMP stars. We leave this as an exercise for later data releases when higher precision abundances become available.

\subsection{Radial velocity variations and binarity}\label{sRV}

Since neutron capture spectral lines are not available in the wavelength regime covered by APOGEE, it is impossible to classify CEMP candidates into the subclasses (CEMP-$s$, CEMP-$r/s$, CEMP-no), however other indicators towards the nature of these stars exist. As mentioned in the introduction, a high binary fraction has been observed for CEMP-$s$ stars. \citet{Lucatello05} inspected the binary frequency among 19 CEMP-$s$ stars by investigating radial velocity (RV) variations with a minimum baseline of $\sim$200 days. They found a binary fraction $\sim 68\%$, higher than the expected detection fraction of $\sim 36\%$ assuming a true binary fraction of 100$\%$, suggesting all CEMP-$s$ are found in binaries. \citet{Starkenburg14} also constrained the binary fraction and periods using tailored Monte Carlo simulations and a maximum-likelihood analysis. Their results reinforce the conclusions made by \citet{Lucatello05}, that a binary fraction of $\sim100\%$ and period distribution $\leq 20,000$ days best represent the RV variations seen within the CEMP-$s$ and classical CH star data groups. Conversely, only two stars in their CEMP-no sample show evidence of short-period radial velocity variations, lightly comparable to observed solar neighbourhood binary fraction ($\sim 45\%$), but highly discrepant to the observed binary properties of the CEMP-$s$ stars. More recently, \citet{Hansen16a}, \citet{Hansen16no}, and \citet{Hansen16s} sought to determine the frequency of binary systems among CEMP stars in a precise and homogeneous manner, with the goal of testing the statistical viability of AGB binary companion mass transfer as a mechanism for carbon enhancement. Nearly a decade of observations on 22 previously identified CEMP-$s$ and CEMP-$r/s$ stars revealed a binary fraction of 82$\pm10\%$. 

\citet{Aoki07}, \citet{Norris13}, \citet{Carollo14}, \citet{Bonifacio15}, \citet{Hansen16a}, and \citet{Yoon16} have demonstrated that CEMP-$s$ stars dominate at [Fe/H]$> - 3$, meaning the higher metallicity CEMP-$s$ stars should be visible to APOGEE, and with such a high binary fraction, binaries should be evident in our sample, which we can explore from the multiple visits of each star carried out by the APOGEE survey.

We examine the RV reported by APOGEE per visit. The APOGEE radial velocities come in two flavours: relative and synthetic. Relative RV's are iteratively determined by comparing the individual visit spectra to a template (the visit spectrum with the highest S/N is used  as the template in the first iteration), shifting the visit spectra to a mean velocity wavelength scale, and creating a combined spectrum to be used as the new template. This process is repeated until the shifted and re-sampled visit spectra converge with the combined spectrum. In order to determine the synthetic RV's, a best-fitting synthetic spectrum is needed. By matching the combined spectrum to a synthetic spectrum from a grid of 538 synthetic spectra, the best-fit synthetic template can be found. From this best-fit synthetic template, the visit spectra are cross-correlated to determine synthetic radial velocities. \citet{Nidever15} suggest the relative velocities should be preferred since they do not depend on the integrity of synthetic library and thus, we adopt relative radial velocity for our analysis.

As a result of the uncertainties in the RV determination, the APOGEE pipeline is unable to directly flag binary systems, however the parameter \texttt{apogeeStar.vscatter} ($\sigma_{v_r}$), the standard deviation of visit spectra RV's relative to the mean, provides the most insight on potential binary status. \citet{Nidever15} investigated the dependence of $\sigma_{v_r}$ as a function of $T_{\rm eff},\  S/N,$ and [Fe/H] and found a peak scatter $\sim$ 100-150m/s, characterizing the intrinsic RV uncertainty in the APOGEE spectra. The authors suggest that if $\sigma_{v_r}\ > 1$ km/s (an order of magnitude larger than the typical uncertainties), then the star is likely a binary. While this assertion is not currently supported in a statistical manner, work is being done to establish a dependable binary identification method (Troup et al., in preparation).

\begin{figure}
\includegraphics[width=\linewidth]{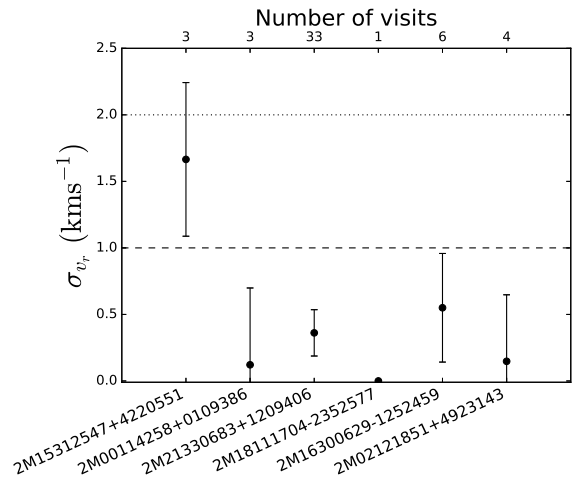}
\caption{Scatter in radial velocity ($\sigma_{v_r}$) for the six Group A stars sorted by $T_{\rm eff}$ (ascending left to right), error bars represent $1/\sqrt{N_{visits}}$. The line at 1 km/s represents our 1$\sigma$ cut for binarity consideration as suggested by the APOGEE team; the dashed line signifying the 2$\sigma$. 2M18111704-2352577 has been included in the figure for completeness, however only one visit makes it impossible to determine variation in the radial velocity.}
\label{RV}
\end{figure}

Figure \ref{RV} summarizes the scatter in RV for the six Group A stars. Only one star in the group, 2M15312547+4220552, has RV scatter greater than 1 km/s. This scatter seen is in agreement with the radial velocity variations seen by \citet{Hansen16s} in their sample of CEMP-$s$ stars ($\sigma_{RV} = 5.6 \pm 5.0$ km/s) and are larger than the mean variations measured by \citet{Starkenburg14} ($\sigma_{RV} = 0.7 \pm 0.8$ km/s). The remaining five candidates show little to no variation, and one star, 2M18111704-2352577, only had one visit. 

This implies a binary fraction for our sample of 16\%. If all our candidates were CEMP-$s$, following the binary fraction by \citet{Hansen16s} of 82$\pm$10\%, we would expect to to find $\sim 5 \pm$ 1 binaries in our sample. Although it is admittedly difficult to comment statistically on such small sample sizes, perhaps our sample is comprised of other CEMP subtypes and/or the presence of binary systems with RV scatter less than 1 km/s like those found in \citet{Starkenburg14} and \citet{Hansen16s}. If our sample includes CEMP-no stars, a lower binary fraction is expected. CEMP-no stars do not exhibit a statistically significant binary fraction \citep{Starkenburg14, Hansen16no}. Following the CEMP-no binary fraction of 17$\pm$9\% found by \citet{Hansen16no}, for a sample comprised only of CEMP-no stars, we would expect $\sim 1 \pm 1$ binaries within our sample, as found, however similar numbers are expected from the binary fraction (16$\pm$4\%) seen in typical field red giants \citep{Carney03}. 

\citep{Spite13, Bonifacio15, Hansen15a, Yoon16} have additionally shown that the nature of these stars may be unlocked through an analysis of the $A$(C) alone. Examining Table \ref{CNOTable}, all of our Group A stars have $7.1 < A$(C)$\lesssim 7.5$, based on the calibrated [C/H] from ASPCAP. This range of $A$(C) is higher than the mid-point of the peaks in the \citet{Yoon16} distribution, placing our stars in the high-C band. This has two primary implications: the Group A stars are more likely to be CEMP-$s$ or CEMP-$r/s$ and they are most likely in binary systems. This second point contradicts our radial velocity analysis, especially for 2M15312547+4220552 which has the lowest $A$(C) in our sample but is also the strongest candidate for a binary system, however the possibility remains our candidates are long period binaries with radial velocity variations that are too small over the duration of the APOGEE observations. A small number of CEMP-no stars and single star systems are found in the high-C band as well, especially at the `intermediate` C-abundances seen in Group A, sustaining the claim Group A is composed of multiple CEMP subtypes. Follow-up optical analysis with access to the CH $G$-band would strengthen the reliability of the carbon abundances derived for the Group A stars and would grant access to neutron-capture spectral lines, allowing for a two direction approach on the classification of these stars.

\subsection{CEMP-no stars in dwarf galaxies}\label{sdSph}

CEMP-no stars have been identified in dwarf galaxies surrounding the Milky Way \citep{Lai11, Honda11, Norris13, Gilmore13, Skuladottir15, Salvadori15, Salgado16}. \citet{Salvadori15} examined the frequency of CEMP-no stars in the dwarf galaxies and concluded CEMP-no stars are expected in all dwarf galaxies with the relative fraction of these stars increasing at lower metallicities, as seen in the Milky Way. However, they also predict the probability to detect these objects scales strongly with the luminosity and the metallicity distribution function (MDF) of the dwarf galaxy. In the more luminous, classical dSph galaxies, CEMP-no stars are expected to be observed at higher metallicities\footnote{The CEMP-no detection likelihood peaks at [Fe/H]$\approx -2$ for a Sculptor-like dSph galaxy.}, on average, with an overall detection probability $P \leq 0.02$. In the ultra-faint systems, a broader MDF and lower luminosity increases the probability of detection to $P \sim 0.1$ with an increased likelihood of observing these stars at lower metallicity.

Congruently, sub-solar $\alpha$-ratios are observed for stars in the Local Group dwarf galaxies as a result of differing star formation histories (e.g. see \citet{Venn04}, \citet{Tolstoy09}, and \citet{FN15}). Highlighted in Section \ref{salpha}, two of the Group A stars, 2M15312547+4220551 and 2M00114258+0109386, have sub-solar $\alpha$-ratios based on their calibrated ASPCAP abundances. Despite large error bars, these low $\alpha$-ratios were confirmed in Section \ref{sMOOGalpha}. These \textit{$\alpha$-challenged} stars (where [$\alpha$/Fe]$< 0.0$) are an uncommon observation at the metallicity of 2M15312547+4220551 and 2M00114258+0109386 ([Fe/H]$\leq -2.0$) indicating these are unique objects more similar to stars in the faint dwarf galaxies than to Galactic field stars. 

Furthermore, \citet{Shetrone01, Shetrone03, Cayrel04, Letarte10, North12}, and \citet{Venn12} have identified sub-solar abundance patterns of odd-Z elements for individual stars in the dwarf galaxies. The ASPCAP DR12 database shows the abundances of the odd-Z elements Na, K, and V for 2M15312547+4220551 and 2M00114258+0109386 are within $1 \sigma$ of scaled solar and do not show any unusual abundance trends. [Mn/Fe] is reported at $-0.27 \pm 0.15 $ and $-0.23 \pm 0.15$ for 2M15312547+4220551 and 2M00114258+0109386, respectively, however Mn is subject to strong HFS effects and must be examined more carefully with optical data. While the positions of these stars do not overlap with the position of any known dSphs in the Local Group, finding new $\alpha$-poor/odd-Z-poor/metal-poor stars that are also C-rich could point towards objects with origins in an accreted dwarf galaxy. High resolution optical analysis is needed to confirm the nature of these seemingly high interest stars.

\subsection{Comparison to Data Release 13}
\begin{table*}
\caption{Comparison of stellar parameters and abundances between DR12 and DR13. The first line for an object shows the DR12 results with the following line containing the DR13 results.  An empty entry means no abundance was determined by ASPCAP.}
\begin{adjustbox}{max width=\textwidth}
\begin{tabular}{cccccccccc}
\hline
ID & $T_{\rm eff}$ & $\log g$ & $\xi$ & [Fe/H] & [C/Fe] & [N/Fe] & [O/Fe] & [$\alpha$/Fe] & Group \\
 & (K) & (dex) & (km s$^{-1}$) & (dex)  & (dex) & (dex) & (dex) & (dex) & Group \\
\hline

DR12 and DR13:\\
2M11584435+5518120 (DR12)  &  4555  &  1.29  &  1.94  &  -1.80  $\pm$  0.06  &  0.74  $\pm$  0.19  &  0.40  $\pm$  0.20  &  0.24  $\pm$  0.10  &  0.28  $\pm$  0.21 & B  \\ 
2M11584435+5518120 (DR13)  &  4502  &  1.34  &  1.67  &  -1.80  $\pm$  0.05  &  0.78  $\pm$  0.12  &  0.59  $\pm$  0.12  &  0.31  $\pm$  0.11  &  0.34  $\pm$  0.16 &  \\ 

2M14571988+1751501 (DR12)  &  4566  &  1.35  &  1.92  &  -1.90  $\pm$  0.06  &  0.78  $\pm$  0.2  &  0.67  $\pm$  0.20  &  0.28  $\pm$  0.10  &  0.19  $\pm$  0.17 & B  \\ 
2M14571988+1751501 (DR13)  &  4516  &  1.37  &  1.67  &  -1.86  $\pm$  0.06  &  0.78  $\pm$  0.18  &  0.83  $\pm$  0.16  &  0.4  $\pm$  0.16  &  0.32  $\pm$  0.11 &  \\ 

2M02121851+4923143 (DR12)  &  4593  &  1.61  &  1.85  &  -1.82  $\pm$  0.06  &  0.82  $\pm$  0.19  &  0.56  $\pm$  0.19  &  0.45  $\pm$  0.09  &  0.42  $\pm$  0.14 & A  \\ 
2M02121851+4923143 (DR13)  &  4517  &  1.65  &  1.65  &  -1.82  $\pm$  0.05  &  0.84  $\pm$  0.17  &  0.67  $\pm$  0.16  &  0.49  $\pm$  0.15  &  0.47  $\pm$  0.17 &  \\ 
\hline

DR12 Only:\\
2M15312547+4220551  &  4068  &  -0.15  &  2.38  &  -2.08  $\pm$  0.06  &  0.75  $\pm$  0.17  &  1.02  $\pm$  0.21  &  -0.15  $\pm$  0.09  &  -0.12  $\pm$  0.45 & A  \\ 
    & ----- & ----- & ----- & ----- & ----- & ----- & ----- & ----- \\
    
2M00114258+0109386  &  4165  &  -0.14  &  2.38  &  -2.18  $\pm$  0.06  &  0.84  $\pm$  0.19  &  1.05  $\pm$  0.22  &  -0.01  $\pm$  0.1  &  0.13  $\pm$  0.12 & A  \\ 
  &  4161  &  1.15  &  1.67  &  -1.67  $\pm$  0.05  &  0.66  $\pm$  0.10  & -----  &  -0.06  $\pm$  0.07  &  -0.15  $\pm$  0.08  \\ 
  
2M21330683+1209406  &  4220  &  0.39  &  2.21  &  -2.01  $\pm$  0.06  &  0.91  $\pm$  0.17  &  0.98  $\pm$  0.19  &  0.29  $\pm$  0.09  &  0.32  $\pm$  0.11 & A  \\ 
  &  4271  &  1.1  &  1.67  &  -1.6  $\pm$  0.05  &  0.92  $\pm$  0.12  &  -----  &  0.19  $\pm$  0.09  &  0.10  $\pm$  0.14  \\ 

2M18111704-2352577  &  4238  &  0.65  &  2.13  &  -1.82  $\pm$  0.06  &  0.76  $\pm$  0.15  &  0.68  $\pm$  0.18  &  0.28  $\pm$  0.08  &  0.27  $\pm$  0.05 & A  \\ 
  &  4298  &  1.2  &  1.67  &  -1.61  $\pm$  0.05  &  0.82  $\pm$  0.10  &  0.91  $\pm$  0.11  &  0.28  $\pm$  0.08  &  0.16  $\pm$  0.07  \\ 

2M16334467-1343201  &  4288  &  0.23  &  2.26  &  -2.1  $\pm$  0.06  &  0.89  $\pm$  0.19  &  1.02  $\pm$  0.2  &  0.19  $\pm$  0.09  &  0.20  $\pm$  0.1 & B  \\ 
  &  4349  &  1.02  &  1.67  &  -1.7  $\pm$  0.05  &  0.86  $\pm$  0.08  &  -----  &  0.08  $\pm$  0.07  &  0.03  $\pm$  0.11  \\ 

2M12473823-0814340  &  4317  &  0.79  &  2.08  &  -2.24  $\pm$  0.06  &  0.75  $\pm$  0.20  &  -0.14  $\pm$  0.22  &  0.87  $\pm$  0.10  &  0.67  $\pm$  0.35 & C  \\ 
  &  4681  &  3.1  &  1.24  &  -0.63  $\pm$  0.03  &  0.12  $\pm$  0.07  &  0.14  $\pm$  0.09  &  0.42  $\pm$  0.07  &  0.36  $\pm$  0.24  \\ 

2M16562103+1002085  &  4376  &  0.93  &  2.05  &  -1.98  $\pm$  0.06  &  0.88  $\pm$  0.18  &  0.95  $\pm$  0.20  &  0.41  $\pm$  0.09  &  0.38  $\pm$  0.06 & B  \\ 
  &  4368  &  1.23  &  1.67  &  -1.76  $\pm$  0.05  &  0.84  $\pm$  0.05  &  0.95  $\pm$  0.06  &  0.25  $\pm$  0.05  &  0.18  $\pm$  0.07  \\ 

2M16300629-1252459  &  4539  &  0.99  &  2.03  &  -1.95  $\pm$  0.06  &  0.97  $\pm$  0.21  &  0.93  $\pm$  0.21  &  0.32  $\pm$  0.10  &  0.31  $\pm$  0.15 & A  \\ 
 &  4537  &  1.22  &  1.67  &  -1.78  $\pm$  0.05  &  0.97  $\pm$  0.09  & -----  &  0.21  $\pm$  0.10  &  0.27  $\pm$  0.21  \\ 

2M16385680+3635073  &  4561  &  1.46  &  1.89  &  -1.84  $\pm$  0.06  &  0.88  $\pm$  0.20  &  0.29  $\pm$  0.20  &  0.23  $\pm$  0.10  &  0.10  $\pm$  0.23 & B  \\ 
  &  4589  &  1.57  &  1.66  &  -1.73  $\pm$  0.05  &  0.95  $\pm$  0.18  &  0.53  $\pm$  0.17  &  0.49  $\pm$  0.16  &  0.21  $\pm$  0.20 \\ 

2M05352696-0510173  &  4591  &  2.34  &  1.64  &  -1.9  $\pm$  0.06  &  0.75  $\pm$  0.22  &  -0.48  $\pm$  0.23  &  0.25  $\pm$  0.11  &  0.37  $\pm$  0.19  & C \\ 
    & ----- & ----- & ----- & ----- & ----- & ----- & ----- & ----- \\

\hline
DR13 Only:\\

2M10331492+2936548  &  4332  &  0.76  &  2.10  &  -1.95  $\pm$  0.06  &  0.62  $\pm$  0.17  &  0.74  $\pm$  0.19  &  0.30  $\pm$  0.09  &  0.22  $\pm$  0.13  \\
 &  4357  &  1.18  &  1.67  &  -1.83  $\pm$  0.05  &  0.73  $\pm$  0.05  &  0.87  $\pm$  0.06  &  0.35  $\pm$  0.05  &  0.31  $\pm$  0.09  \\

2M16235550-1323093  &  4614  &  1.34  &  1.93  &  -1.97  $\pm$  0.06  &  0.85  $\pm$  0.24  &  -0.32  $\pm$  0.23  &  0.34  $\pm$  0.11  &  0.58  $\pm$  0.24  \\ 
  &  4538  &  1.19  &  1.67  &  -1.90  $\pm$  0.06  &  0.71  $\pm$  0.12  &  0.07  $\pm$  0.11  &  0.68  $\pm$  0.12  &  0.47  $\pm$  0.20  \\ 

2M10095022+0159202  &  4827  &  1.61  &  1.85  &  -1.84  $\pm$  0.07  &  0.71  $\pm$  0.28  &  0.28  $\pm$  0.25  &  0.08  $\pm$  0.13  &  0.27  $\pm$  0.22  \\
 &  4569  &  1.31  &  1.67  &  -2.07  $\pm$  0.06  &  0.76  $\pm$  0.22  &  0.41  $\pm$  0.18  &  0.50  $\pm$  0.20  &  0.41  $\pm$  0.24  \\  

2M21300316+1213286  &  4834  &  1.27  &  1.95  &  -2.07  $\pm$  0.07  &  0.71  $\pm$  0.28  &  0.58  $\pm$  0.25  &  0.58  $\pm$  0.12  &  0.35  $\pm$  0.21  \\
  &  4573  &  0.89  &  1.67  &  -2.24  $\pm$  0.06  &  0.80  $\pm$  0.23  &  0.63  $\pm$  0.18  &  0.36  $\pm$  0.21  &  0.41  $\pm$  0.45  \\  

2M14055319+5227233  &  4862  &  1.82  &  1.79  &  -1.97  $\pm$  0.07  &  0.30  $\pm$  0.29  &  0.64  $\pm$  0.25  &  0.51  $\pm$  0.13  &  0.38  $\pm$  0.14  \\ 
 &  4506  &  1.17  &  1.67  &  -2.22  $\pm$  0.06  &  0.80  $\pm$  0.19  &  0.58  $\pm$  0.16  &  0.57  $\pm$  0.18  &  0.40  $\pm$  0.22  \\ 

2M18225322+0113105  &  4915  &  3.18  &  1.4  &  -1.78  $\pm$  0.06  &  0.83  $\pm$  0.25  &  -0.15  $\pm$  0.22  &  0.50  $\pm$  0.12  &  0.42  $\pm$  0.17  \\
 &  4512  &  2.43  &  1.54  &  -2.11  $\pm$  0.06  &  0.75  $\pm$  0.19  &  0.01  $\pm$  0.16  &  0.18  $\pm$  0.17  &  0.32  $\pm$  0.26  \\  

2M13205580+1231196  &  5033  &  2.04  &  1.73  &  -2.03  $\pm$  0.07  &  0.71  $\pm$  0.34  &  0.88  $\pm$  0.26  &  0.55  $\pm$  0.14  &  0.43  $\pm$  0.11  \\
  &  4510  &  1.06  &  1.67  &  -2.37  $\pm$  0.06  &  0.81  $\pm$  0.22  &  0.48  $\pm$  0.18  &  0.56  $\pm$  0.20  &  0.59  $\pm$  0.26  \\  

2M15144943-0141354  &  5267  &  2.66  &  1.55  &  -1.68  $\pm$  0.06  &  0.68  $\pm$  0.33  &  0.56  $\pm$  0.24  &  0.34  $\pm$  0.14  &  0.05  $\pm$  0.32  \\
  &  4457  &  1.21  &  1.67  &  -2.22  $\pm$  0.06  &  0.76  $\pm$  0.10  &  0.12  $\pm$  0.10  &  0.50  $\pm$  0.11  &  0.32  $\pm$  0.31  \\  

2M12221630+1407311  &  5315  &  2.34  &  1.64  &  -----  &  -----  &  -----  &  -----  &  -----  \\ 
  &  4596  &  1.18  &  1.67  &  -1.98  $\pm$  0.06  &  0.76  $\pm$  0.19  &  0.52  $\pm$  0.16  &  -----  &  0.28  $\pm$  0.33  \\ 

2M07573998+4038029  &  5351  &  2.76  &  1.52  &  -----  &  -----  &  -----  &  -----  &  -----  \\ 
  &  4466  &  1.15  &  1.67  &  -2.37  $\pm$  0.06  &  0.84  $\pm$  0.02  &  0.42  $\pm$  0.02  &  0.46  $\pm$  0.04  &  0.28  $\pm$  0.12  \\ 

2M17172331+4256556  &  5448  &  2.93  &  1.47  &  -----  &  -----  &  -----  &  ----- &  -----  \\
  &  4511  &  1.36  &  1.67  &  -2.27  $\pm$  0.06  &  0.95  $\pm$  0.21  &  0.20  $\pm$  0.17  &  0.09  $\pm$  0.19  &  0.36  $\pm$  0.37  \\  

2M17175037+4313460  &  5604  &  3.21  &  1.39  &  -----  &  -----  &  -----  &  -----  &  -----  \\
  &  4597  &  1.42  &  1.66  &  -2.28  $\pm$  0.06  &  0.84  $\pm$  0.22  &  0.47  $\pm$  0.18  &  0.57  $\pm$  0.21  &  0.57  $\pm$  0.30  \\  

2M11464977+2746407  &  6180  &  3.79  &  1.23  &  -----  &  -----  & ----- &  -----  &  -----  \\
  &  4435  &  0.77  &  1.66  &  -2.35  $\pm$  0.06  &  0.79  $\pm$  0.14  &  0.55  $\pm$  0.12  &  0.60  $\pm$  0.14  &  0.16  $\pm$  0.32  \\  

2M14514157+1623464    & ----- & ----- & ----- & ----- & ----- & ----- & ----- & ----- \\
  &  4381  &  0.72  &  1.66  &  -2.25  $\pm$  0.07  &  0.75  $\pm$  0.25  &  0.43  $\pm$  0.19  &  0.72  $\pm$  0.19  &  0.44  $\pm$  0.21  \\   
  
2M16563200+1024306   & ----- & ----- & ----- & ----- & ----- & ----- & ----- & ----- \\ 
&  4542  &  1.27  &  1.67  &  -2.15  $\pm$  0.06  &  0.95  $\pm$  0.07  &  0.91  $\pm$  0.08  &  0.40  $\pm$  0.09  &  0.38  $\pm$  0.11  \\ 

2M13091378-0230485  & ----- & ----- & ----- & ----- & ----- & ----- & ----- & ----- \\
  &  4581  &  1.34  &  1.67  &  -2.34  $\pm$  0.07  &  0.86  $\pm$  0.25  &  0.68  $\pm$  0.19  &  0.62  $\pm$  0.23  &  0.53  $\pm$  0.23  \\ 
\hline
\end{tabular}
\end{adjustbox}
\label{DR12vDR13}
\end{table*}

The bulk of this analysis has focused on the data available through DR12, however the recent release of DR13 \citep{Albareti16} allows us to internally compare the ASPCAP results. DR13 uses the same data from DR12 but with improved data reduction techniques, an updated linelist, new spectral grids for dwarfs and giants, as well as an updated calibration of abundances \citep[][others in prep.]{GarciaPerez16}. All cool CEMP candidates ($T_{\rm eff} < 4600$K, [Fe/H]$< -1.8$, and [C/Fe]$>+0.7$) from DR13 were selected and cross-matched with our initial sample of 13 CEMP candidates from DR12  in Table \ref{DR12vDR13}. Three stars, 2M02121851+4923143 (Group A), 2M11584435+5518120 (Group B), and 2M14571988+1751501 (Group B) were identified by ASPCAP as CEMP candidates in both DR12 and DR13. The stellar parameters, as well as the abundances of Fe and C for these three stars are in excellent agreement between DR12 and DR13. The N and O abundances are additionally consistent to within the error bars, however a small degree of scatter in the abundances is seen for the two Group B stars. Since our Group B stars were affected by persistence in DR12 and that DR13 sought to address the persistence problem, this variance is likely the result of the new analysis. Examining the individual visit spectra from DR13, we note the persistence appears to be corrected for three of the six Group B targets, including 2M11584435+5518120 and 2M14571988+1751501. The higher data quality in DR13 makes these two stars prime candidates for follow up analysis.

The 10 remaining DR12 CEMP candidates between Groups A, B, and C are not identified as CEMP in DR13. With the exception of the two coolest Group A stars, 2M15312547+4220551 and 2M00114258+0109386, the remaining Group A and Group B candidates remain classified as carbon-enhanced in DR13, but their DR13 metallicities are above the [Fe/H]$\leq -1.8$ limit imposed for CEMP. Offsets in metallicity on the order of 0.2 dex appears typical between DR12 and DR13 for this selected sample. Similar offsets in [O/Fe] are observed suggesting the [O/H] abundance between DR12 and DR13 is robust, however the the derivation of the N abundance is suffering from pipeline limitations in DR13. Additionally, a majority of these objects show a striking disparity in $\log g$ between the two data releases. Not only do these differences suggest different evolutionary states for some objects, but a change in $\log g$ on the order of $\pm 1.0$ dex could result in abundance variations for all species up to $\pm 1.0$ dex as well (see Table \ref{SysErrs}). These discrepancies between DR12 and DR13 assert the need for a star-by-star analysis of the APOGEE CEMP candidates.

We identify 16 new CEMP candidates in DR13 that were not included in our sample from DR12. Eight of these stars have $T_{\rm eff} > 4600K$  or fail to meet the metallicity or carbonicity requirements for the CEMP classification when adopting their DR12 stellar parameters and abundances. Abundances and/or stellar parameters were not reported in DR12 for the remaining eight stars identified as CEMP by DR13. A future study into the nature of the DR13 CEMP candidates is warranted.

\section{Conclusions}
This work serves as the first exploration of CEMP stars using $H$-band spectra. Investigation of the DR12 APOGEE ASPCAP database for CEMP candidates has resulted in 13 stars with [C/Fe]$>0.7$ and [Fe/H]$<-1.8$. Within this sample, six stars (Group A) were identified to have "clean" spectra, free of flat fielding and persistence, and well matched synthetic spectra. The remaining seven stars showed moderate to strong evidence of persistence and/or poorly matched synthetic spectra, decreasing the likelihood that ASPCAP can identify the CEMP signatures. Temperature effects, abundance upper-limits, molecular blends, and other systematics have additionally complicated the accuracy of the ASPCAP abundances. These stars serve as a caution when inspecting the APOGEE database for metal-poor and chemically peculiar stars. A model atmospheres analysis and \textsc{moog} spectrum syntheses were performed for each of the Group A candidates, confirming their CEMP classification, but raising further questions on their nitrogen and $\alpha$-element enhancements and the reliability of the ASPCAP abundances at low metallicity.

The extent of the ASPCAP data has additionally allowed for a preliminary study into the nature of these stars. Low $\alpha$-abundances are observed for two Group A stars, encouraging the possibility their origins are in an accreted dwarf galaxy, and a binary fraction of 16\% (one star) was determined for Group A by exploring the radial velocity variations between individual visits. A higher binary fraction was expected if the dataset was comprised solely of CEMP-$s$ stars \citep{Carney03, Starkenburg14, Hansen16no}, indicating our sample may contain a more diverse and revealing group of stars in the framework of Galactic chemical evolution. Follow up observations of these objects are required to determine the $s-$ and $r-$ process abundances necessary for further sub-classification. An inceptive look into DR13 has additionally uncovered up to 16 CEMP candidates worth investigating with similar methods presented here for the DR12 spectra.

The depth of information accessible through APOGEE spectra exceeds what is currently available in the ASPCAP database. Careful and targeted analyses of this large database have and will continue to yield new insight into both fundamental and unique astrophysical processes, Galactic structure, chemistry, and evolution.

\section*{Acknowledgements}
We would like to thank P. Bonifacio, N. Martin, and the Stars Group at University of Victoria for stimulating and constructive discussions as well as the anonymous referee for their supportive comments that have improved this paper. Szabolcs M{\'e}sz{\'a}ros has been supported by the Premium Postdoctoral Research Program of the Hungarian Academy of Sciences, and by the Hungarian NKFI Grants K-119517 of the Hungarian National Research, Development and Innovation Office. Funding for SDSS-III has been provided by the Alfred P. Sloan Foundation, the Participating Institutions, the National Science Foundation, and the U.S. Department of Energy Office of Science. The SDSS-III web site is http://www.sdss3.org/. SDSS-III is managed by the Astrophysical Research Consortium for the Participating Institutions of the SDSS-III Collaboration including the University of Arizona, the Brazilian Participation Group, Brookhaven National Laboratory, Carnegie Mellon University, University of Florida, the French Participation Group, the German Participation Group, Harvard University, the Instituto de Astrofisica de Canarias, the Michigan State/Notre Dame/JINA Participation Group, Johns Hopkins University, Lawrence Berkeley National Laboratory, Max Planck Institute for Astrophysics, Max Planck Institute for Extraterrestrial Physics, New Mexico State University, New York University, Ohio State University, Pennsylvania State University, University of Portsmouth, Princeton University, the Spanish Participation Group, University of Tokyo, University of Utah, Vanderbilt University, University of Virginia, University of Washington, and Yale University.

\bibliographystyle{mnras}
\bibliography{Refs}

\appendix
\section{Line Lists}
\begin{table*}
\caption{Atomic lines and derived log abundances for the $\alpha$-elements Mg, Si, S, and Ca}

\begin{tabular}{lccccccccc}
\hline
Element & Wavelength (\AA) & $\chi$(eV) & $\log \ gf$ & 2M153... & 2M001... & 2M213... & 2M181... & 2M163... & 2M021... \\
\hline

\ion{Mg}{i}&	15745.017	&	5.931	&	-0.262	&	-----	&	-----	&	-----	&	-----	&	-----	&	5.93	\\
	&	15753.189	&	5.932	&	-0.388	&	5.0	&	4.8	&	5.4	&	5.8	&	5.6	&	5.9	\\
	&	15770.055	&	5.933	&	-0.387	&	4.8	&	5.0	&	5.4	&	5.7	&	5.6	&	5.9	\\																		
\ion{Si}{i}	&	15893.2	&	7.125	&	-2.49	&	-----	&	5.2	&	-----	&	5.4	&	5.2	&	-----	\\
	&	15963.99	&	7.029	&	-2.319	&	-----	&	4.7	&	-----	&	5.9	&	5.7	&	-----	\\
	&	16064.4	&	5.954	&	-0.566	&	-----	&	4.7	&	5.7	&	5.9	&	5.7	&	6.1	\\
	&	16099.18	&	5.964	&	-0.168	&	-----	&	-----	&	5.6	&	5.8	&	5.6	&	5.9	\\
	&	16168	&	5.954	&	-0.937	&	$< $5.5	&	4.9	&	5.8	&	5.9	&	5.7	&	6.0	\\
	&	16685.33	&	5.94	&	-0.14	&	5.0	&	5.2	&	5.5	&	5.7	&	5.6	&	5.9	\\																		
\ion{S}{i}	&	15426.48&	8.7	&	-0.238	&	-----	&	-----	&	-----	&	-----	&	-----	&	$< $5.6	\\
\ion{Ca}{ii}	&	16201.5 	&	4.535	&	0.092	&	----- &	-----	&	-----	&	-----	&	-----	&	$<$4.7	\\
\hline
\end{tabular}
\label{AlphaLinelist}
\end{table*}

\begin{table*}
\caption{Molecular features and log abundances used to derive C, N, and O.}
\begin{adjustbox}{max height=11 cm}
\begin{tabular}{lccccccc}
\hline
Element & Wavelength Interval (\AA) & 2M153... & 2M001... & 2M213... & 2M181... & 2M163... & 2M021... \\
\hline

C from atomic \ion{C}{i} line 	&	16895.03	&	7.7	&	7.6	&	7.5	&	7.5	&	7.7	&	7.4	\\
O from OH lines	&	15283-15287	&	-----	&	-----	&	-----	&	7.9	&	-----	&	7.7	\\
	&	15560-15570	&	-----	&	-----	&	-----	&	8.0	&	-----	&	-----	\\
	&	15571-15578	&	-----	&	-----	&	7.4	&	-----	&	-----	&	7.5	\\
	&	15721	&	-----	&	-----	&	-----	&	8.2	&	-----	&	7.9	\\
	&	15724	&	-----	&	-----	&	-----	&	7.7	&	-----	&	7.1	\\
	&	15731	&	-----	&	-----	&	-----	&	8.1	&	-----	&	7.9	\\
	&	15735	&	-----	&	-----	&	-----	&	7.9	&	-----	&	7.6	\\
	&	15760	&	-----	&	-----	&	-----	&	-----	&	-----	&	7.3	\\
	&	15777-15780	&	7.0	&	7.2	&	-----	&	7.2	&	7.4	&	7.2	\\
	&	15783	&	-----	&	-----	&	-----	&	-----	&	-----	&	7.4	\\
	&	15895-15899	&	-----	&	-----	&	-----	&	7.7	&	8.1	&	7.6	\\
	&	15913-15919	&	6.8	&	-----	&	-----	&	7.9	&	7.6	&	7.6	\\
	&	16057	&	7.0	&	-----	&	7.6	&	-----	&	7.6	&	7.4	\\
	&	16060	&	7.2	&	-----	&	7.6	&	-----	&	-----	&	7.4	\\
	&	16069	&	6.6	&	-----	&	-----	&	-----	&	-----	&	7.6	\\
	&	16074	&	7.0	&	-----	&	-----	&	-----	&	-----	&	7.4	\\
	&	16079	&	-----	&	-----	&	-----	&	-----	&	-----	&	7.6	\\
	&	16184	&	-----	&	-----	&	7.9	&	7.9	&	-----	&	-----	\\
	&	16187-16200	&	-----	&	-----	&	-----	&	7.9	&	-----	&	7.3	\\
	&	16208	&	-----	&	-----	&	7.6	&	7.6	&	7.7	&	7.5	\\
	&	16224	&	-----	&	-----	&	7.6	&	7.6	&	-----	&	7.6	\\
	&	16235-16238	&	-----	&	-----	&	-----	&	-----	&	-----	&	7.7	\\
	&	16270	&	-----	&	-----	&	-----	&	-----	&	-----	&	7.3	\\
	&	16362-16384	&	-----	&	6.8	&	-----	&	-----	&	-----	&	7.0	\\
	&	16523-16533	&	-----	&	6.8	&	-----	&	-----	&	-----	&	7.3	\\
	&	16616-16650	&	-----	&	7.1	&	7.3	&	7.3	&	-----	&	7.2	\\
	&	16653-16677	&	-----	&	7.2	&	-----	&	-----	&	-----	&	7.5	\\
	&	16706-16716	&	-----	&	7.2	&	-----	&	-----	&	-----	&	7.5	\\
	&	16731-16739	&	-----	&	-----	&	-----	&	-----	&	-----	&	7.4	\\
	&	16758-16761	&	-----	&	-----	&	-----	&	-----	&	7.8	&	-----	\\
	&	16763-16766	&	-----	&	-----	&	-----	&	-----	&	7.4	&	7.1	\\
	&	16770	&	-----	&	6.8	&	-----	&	-----	&	8.0	&	7.3	\\
	&	16836-16847	&	7.2	&	7.2	&	-----	&	-----	&	-----	&	7.3	\\
	&	16875-16880	&	-----	&	7.0	&	-----	&	-----	&	-----	&	7.4	\\
	&	16888-16893	&	6.6	&	7.0	&	-----	&	-----	&	-----	&	7.3	\\
	&	16897-16911	&	-----	&	-----	&	-----	&	-----	&	7.5	&	7.5	\\

N from CN lines	&	15193	&	-----	&	-----	&	7.4	&	6.9	&	7.4	&	-----	\\
	&	15186-15206	&	-----	&	6.5	&	-----	&	-----	&	-----	&	6.6	\\
	&	15220	&	-----	&	6.6	&	7.3	&	6.9	&	7.3	&	6.7	\\
	&	15226	&	-----	&	6.8	&	-----	&	-----	&	7.3	&	6.9	\\
	&	15229-15234	&	-----	&	6.7	&	-----	&	-----	&	7.3	&	6.8	\\
	&	15252-15261	&	-----	&	6.6	&	7.0	&	6.8	&	7.0	&	6.7	\\
	&	15263-15270	&	-----	&	-----	&	7.3	&	6.9	&	-----	&	-----	\\
	&	15301	&	-----	&	6.5	&	7.4	&	7.0	&	7.3	&	6.7	\\
	&	15320-15336	&	-----	&	6.6	&	7.2	&	6.9	&	7.2	&	6.7	\\
	&	15365-15372	&	6.3	&	-----	&	7.4	&	7.0	&	7.4	&	-----	\\
	&	15377	&	6.2	&	-----	&	7.2	&	-----	&	7.2	&	-----	\\
	&	15399	&	6.9	&	-----	&	7.2	&	-----	&	7.2	&	-----	\\
	&	15401	&	6.5	&	-----	&	7.1	&	6.9	&	7.1	&	6.7	\\
	&	15404	&	6.5	&	-----	&	7.3	&	-----	&	7.3	&	-----	\\
	&	15414	&	6.7	&	-----	&	7.4	&	-----	&	7.4	&	-----	\\
	&	15426-15432	&	6.2	&	-----	&	7.2	&	-----	&	7.2	&	-----	\\
	&	15438	&	6.5	&	6.2	&	6.9	&	7.0	&	6.9	&	6.6	\\
	&	15440	&	6.5	&	6.2	&	6.9	&	-----	&	6.9	&	6.6	\\
	&	15443	&	6.5	&	-----	&	7.4	&	-----	&	7.4	&	-----	\\
	&	15451	&	-----	&	-----	&	7.2	&	-----	&	7.2	&	-----	\\
	&	15453-15459	&	-----	&	-----	&	6.8	&	-----	&	6.8	&	-----	\\
	&	15480-15484	&	-----	&	-----	&	6.8	&	-----	&	6.8	&	-----	\\
	&	15486	&	-----	&	-----	&	7.2	&	-----	&	7.2	&	-----	\\
	&	15512	&	-----	&	6.4	&	7.4	&	-----	&	7.4	&	6.8	\\
	&	15534	&	-----	&	-----	&	7.1	&	-----	&	7.1	&	-----	\\
	&	15543-15553	&	-----	&	-----	&	7.2	&	-----	&	7.2	&	-----	\\
	&	15598	&	-----	&	6.5	&	-----	&	6.9	&	-----	&	6.8	\\
	&	15617	&	-----	&	6.6	&	-----	&		&	-----	&	6.8	\\
	&	15669	&	-----	&	6.4	&	-----	&	7.0	&	-----	&	6.8	\\
	&	15699-15713	&	-----	&	-----	&	-----	&	7.0	&	7.1	&	-----	\\
	&	15674-15687	&	-----	&	6.4	&	-----	&	-----	&	7.2	&	6.7	\\
	&	15727-15731	&	-----	&	-----	&	-----	&	6.9	&	-----	&	-----	\\
	&	15738-15743	&	-----	&	-----	&	-----	&	6.9	&	-----	&	-----	\\
	&	15771-15787	&	-----	&	6.6	&	-----	&	6.4	&	-----	&	-----	\\
	&	15896-15899	&	-----	&	6.2	&	-----	&	-----	&	-----	&	6.4	\\
	&	15907-15927	&	-----	&	6.4	&	-----	&	-----	&	-----	&	6.7	\\
    \hline
\end{tabular}
\end{adjustbox}
\label{CNOLinelist}
\end{table*}

\bsp	
\label{lastpage}
\end{document}